\documentclass[twocolumn]{aastex701} 

\usepackage{graphicx} 
\usepackage{float}
\usepackage[section]{placeins}
\usepackage{amsmath}
\usepackage{diagbox}
\usepackage{booktabs}
\usepackage{comment}

\begin{document}
\title{A Leptonic Interpretation of the UHE Gamma-ray Emission from V4641~Sgr}

\author{Su-Yu Wan}
\affiliation{School of Astronomy and Space Science, Nanjing University, Nanjing 210023, China}
\affiliation{Key laboratory of Modern Astronomy and Astrophysics(Nanjing University),Ministry of Education,Nanjing 210023, People's Republic of China}
\email{sywan2002@smail.nju.edu.cn}

\author{Jie-shuang Wang}
\affiliation{Max Planck Institute for Plasma Physics, Boltzmannstra{\ss}e 2, D-85748 Garching, Germany }
\email{jiesh.wang@gmail.com}

\author{Ruo-Yu Liu}
\affiliation{School of Astronomy and Space Science, Nanjing University, Nanjing 210023, China}
\affiliation{Key laboratory of Modern Astronomy and Astrophysics(Nanjing University),Ministry of Education,Nanjing 210023, People's Republic of China}
\affiliation{Tianfu Cosmic Ray Research Center,
 Chengdu 610000, Sichuan, China}
\email{ryliu@nju.edu.cn}

\correspondingauthor{Ruo-Yu Liu}
\email{ryliu@nju.edu.cn}

\begin{abstract}

Recently, the microquasar V4641~Sgr and its surrounding is detected at TeV-PeV gamma-ray band. Interestingly, the spectrum follows a power-law function continuing up to 0.8\,PeV as reported by LHAASO, and the morphology of the emission appears a puzzling elongated structure. In this work, we propose that the elongated UHE emission from V4641~Sgr could  originate from the inverse Compton radiation of electrons with a very hard spectrum, which may result from shear acceleration mechanism in the jets driven by V4641~Sgr. We also calculate the corresponding X-ray synchrotron emission from the same electron population, predicting the potential range of non-thermal X-ray flux of the source. The recent observation by XRISM toward the central part of the UHE source could pose a constraint on the model parameters. In the future, a full coverage of the source by sensitive X-ray instrument and high-resolution TeV-PeV gamma-ray instrument may provide a critical test of the model.      

\end{abstract}


\section{Introduction} \label{sec:intro}

Microquasars are a subclass of X-ray binary systems \citep{bosch2009understanding}, composed of an accreting compact object and a donor companion star. In the process of mass transfer, structures like relativistic jets could be formed. Microquasars have been found in our own Milky Way, so their jets could be spatially resolved by TeV gamma-ray instrument with angular resolution of $\sim 0.1^\circ$. 
They may serve as natural laboratories for studying particle acceleration in relativistic jets, and have been suggested as main sources of cosmic rays at and beyond the ``knee'' in the cosmic ray spectrum \citep{lhaaso2024ultrahigh, Wang:ApJL:2025, Zhang2025}. 
Recently, five microquasars have been detected at UHE gamma-ray band by the Large High Altitude Air Shower Observatory (LHAASO)\citep{lhaaso2024ultrahigh}, High Altitude Water Cherenkov Observatory (HAWC)\citep{alfaro2024ultra}, and High Energy Stereoscopic System (H.E.S.S.)\citep{hess2024acceleration}. 
Among these detected microquasars, V4641~Sgr, a very active low-mass X-ray binary located at 6.2\,kpc away from Earth \citep{macdonald2014black,gandhi2019gaia}, stands out for the gamma-ray spectrum which continues up to 0.8 PeV following a power-law distribution of slope $-2.67$ above 40\,TeV \citep{lhaaso2024ultrahigh}. Such high-energy photons indicate the presence of efficient particle acceleration in this system. The morphology of the UHE gamma-ray source shows an elongated morphology \citep{lhaaso2024ultrahigh,alfaro2024ultra}, and the physical size along the extension direction reaches $\sim 100\,$pc. The morphological analysis shows that the emission may be also fitted by two ``lobes''. \citealt{alfaro2024ultra} suggested that the emission of V4641~Sgr is due to hadronic radiation of protons accelerated by jet termination shocks. The source is also suggested to be produced by escaping protons \citep{ohira2024very, neronov2025multimessenger}, where the elongated morphology could be ascribed to the topology of the local Galactic magnetic field \citep{neronov2025multimessenger}.

While these studies suggested hadronic origin of the UHE gamma-ray emission, the leptonic scenario, in
 which the UHE photons are generated by inverse Compton (IC) radiation of high-energy electrons, has not been investigated in depth. Indeed, the hadronic scenario has a few advantages over the leptonic scenario in explaining the spectrum and the large extension of the source. First, protons do not suffer strong radiative cooling, so it would be easier for protons to be accelerated to PeV level and diffuse over a large distance. Second, the gamma-ray spectrum of hadronic radiation approximately follows that of protons, whereas the IC radiation of electrons at UHE band will suffer the Klein-Nishina (KN) effect and generally a softening is expected. However, the hadronic scenario probably suffers energetic problem. The luminosity of the source at $\sim$ 100\,TeV is measured to be $L_{\rm UHE.obs}\approx 2\times 10^{34}$ erg/s \citep{lhaaso2024ultrahigh}. With the proton energy loss timescale in $pp$ collisions to be $t_{pp} \simeq 1.3\times 10^{15}(n_{\rm H}/1\, \rm cm^{-3})^{-1}\,s$ with $n_{\rm H}$ being the hydrogen density of ambient interstellar medium (ISM), the total energy of PeV protons in the system need be $W_{p, \rm PeV}\approx  3L_{\rm UHE, obs}t_{pp}\approx 10^{50}(n_{\rm H}/1\,\rm cm^{-3})^{-1}$\,erg. The prefactor 3 comes from the fact that about 1/3 of the energy lost by protons goes into neutral pions which then decay into gamma rays. On the other hand, these protons need be confined within the source which has an extension of $R\sim 100\,$pc. Assuming the 
 diffusion (i.e., the slowest diffusion) in typical interstellar magnetic field of $1\mu\,$G for PeV protons, we may estimate $D_{\rm ISM, PeV}\sim 3\times 10^{28}\,\rm cm^2/s$, and obtain the confinement timescale of PeV protons within the source to be  $t_{\rm conf, PeV}=R^2/4D_{\rm ISM, PeV}\sim 10^{12}\,$s or approximately 30\,kyr. That is to say, PeV protons need be injected from the microquasar at an average luminosity of $W_{p,\rm PeV}/t_{\rm conf, PeV}\sim 10^{38}\,$erg/s\footnote{Here we employ a typical ISM density $n_{\rm H}=1\,\rm cm^{-3}$ for a conservative estimation.} during the past 30\,kyr. The required total proton luminosity above 1\,GeV is likely one order of magnitude higher, and even exceeds the Eddington luminosity of the BH.

Note that, at $b=-4.8^\circ$ and 6.2\,kpc from the Earth, V4641~Sgr is located at about 500\,pc above the Galactic plane, so the density of its ambient medium is likely to be lower than the typical ISM density of $n_{\rm H}=1\,\rm cm^{-3}$, unless dense gamma-ray emitting regions around the source are reported. {Given the extreme and unrealistic requirements from the hadronic scenario (at least for a pure hadronic scenario) in such cases, it is of great necessity to explore} whether the UHE emission associated with V4641~Sgr can be explained under the leptonic scenario. 
We envisage the following picture: the frequent outbursts of V4641~Sgr, have contributed to the formation a pair of quasi-continuous jets, and the period for the bursts (or flare-like activities) is around one to two years \citep{markwardt1999variable, hjellming2000light, buxton2003optical,  swank2004x, uemura2004outburst, cackett2007swift, yamaoka2008swift, yamaoka2010rxte, tachibana2014maxi, yoshii2015maxi, tetarenko2016watchdog}. Consequently, instead of being terminated at relatively small distance, the two mildly-relativistic jets have propagated to $\sim 100$\,pc scale away from the central black hole. A velocity shear across the jet or spine-layer structure may form because of the interaction with the ambient medium\citep{WangJS2023, WangJS2024}. A `distributed' acceleration of electrons along the jet may then occur via the shear acceleration (SHA) mechanism \citep{berezhko1981kinetic,earl1988cosmic, rieger2006microscopic, liu2017particle, webb2018particle, Webb2019,rieger2019introduction, WangJS2024}, which help explain the large extension of the source. Such a mechanism has been suggested to explain the kpc-scale TeV jet seen in Centaurus A by H.E.S.S. \citep{HESS2020} and extended X-ray emissions in kpc-Mpc scale jets launched from some radio galaxies \citep{liu2017particle, wang2021particle, He2023}. A hard particle spectrum is also expected in the SHA mechanism, which may compensate the spectrum softening caused by the KN effect and produce a hard spectrum in a power-law distribution approximately up to PeV energy \citep{Liu2021crab}. Finally, as a supplementary note, we argue that the intermittent outbursts would not pose significant influences on the observed gamma-ray source's morphology. Assuming V4641~Sgr underwent one major outburst per year with a mildly-relativistic jet velocity ($\sim0.7c$), the gap in the jet due to the outburst is $\delta L=0.7c  \times 1\, \rm yr \approx 0.2\,$pc. For a total jet length of $\sim 100\,$pc, a few years' quiescence of outburst events would not cause significant variable luminosity. On the other hand, considering the distance of V4641~Sgr to be $d=6.2\,$kpc, the angular size of the gap is around $\Omega = \Delta L/d\simeq 7''$. Such a small gap cannot be resolved by extensive air shower arrays such as LHAASO and HAWC which have angular resolution of  $0.2^\circ-0.3^\circ$ nor by imaging air Cherenkov telescopes such as HESS with angular resolution of $\sim 0.05^\circ$. Moreover, the spatial diffusion of accelerated particles along the jet might have already smeared out the discontinuity. Taking those factors into account, an elongated and continuous structure could be expected.

The rest of the paper is organized as follows: in Section \ref{sec:spec}, we introduce our method and the model setup; in Section \ref{sec:NonThermal}, we show our fitting results to the gamma-ray data and constraints from X-ray observations; we discuss our model in Section \ref{sec:Discus} and give conclusions in Section \ref{sec:Conclusions}.

\section{Model and Methods} \label{sec:spec}

\subsection{The jet model and electron spectra}\label{sec: jetmodel_spec}

Considering the interactions between a jet and its surrounding medium, deceleration of the jet may occur in the boundary region, while the central velocity remains nearly unchanged. Hence a velocity gradient along the transverse direction of the jet axis may form. Such jets with velocity shear have been reported in some high-resolution radio observations
\citep{nagai2014limb,gabuzda2014parsec,laing2014systematic}. It has been suggested that particles can be scattered off magnetic turbulence frozen in different velocity layers throughout the jet, gaining energies after sufficient times of scatterings \citep{rieger2006microscopic}, which is a kind of second-order Fermi particle acceleration mechanism. The acceleration rate is related to the shear velocity profile. In this work we assume that the jet is composed of a central spine with constant speed $\beta_0$ (in unit of the speed of light $c$) and a velocity-shear boundary, and for simplicity a linear profile is adopted as
\begin{equation}
\beta(r)=
\begin{cases}
\beta_0& 0 \leq r<\left(1-\eta\right)R_{\rm jet}\,, \\
\frac{\beta_0}{\eta}\left(1-\frac{r}{R_{\rm jet}}\right) & \left(1-\eta\right)R_{\rm jet}\leq r \leq R_{\rm jet}\,.
\end{cases}
\label{e1}
\end{equation}
Here $\eta$ is a dimensionless parameter, indicating the ratio between the size of the velocity-shear boundary($R_{\rm SHA}$) and the radius of the jet ($R_{\rm jet}$). With the assumed velocity profile, velocity gradients ($\partial \beta/\partial r$) in the two flow regions can also be readily calculated.

SHA generally has a higher acceleration rate when accelerating higher-energy particles. Therefore, at low energies it requires `seed particles' pre-accelerated by other mechanisms, since otherwise the overall acceleration would become inefficient.
It has been demonstrated that \citep{liu2017particle} the stochastic acceleration (STA) accompanies with the SHA, and can naturally serve as pre-acceleration process although itself is not an efficient mechanism in weakly magnetized jets. Both STA and SHA can be regarded as `diffusion' processes in the momentum phase space of particles, and acceleration process can be described with the Fokker-Planck-type equation \citep{rieger2007fermi, liu2017particle}. For simplicity, we assume the diffusion is isotropic. Here the equation is written in the energy form, which reflects the energy distribution of accelerated particles more directly: 
\begin{equation}
\begin{aligned}
    &\frac{\partial n\left(\gamma,t\right)}{\partial t}=\frac{1}{2}\frac{\partial}{\partial \gamma}\left[\langle \frac{\Delta \gamma^2}{\Delta t} \rangle \frac{\partial n\left(\gamma,t\right)}{\partial \gamma}\right]\\&-\frac{\partial}{\partial \gamma}\left[\left(\langle \frac{\Delta \gamma}{\Delta t} \rangle-\frac{1}{2}\frac{\partial}{\partial \gamma}\langle \frac{\Delta \gamma^2}{\Delta t} \rangle+\left \langle \dot{\gamma_c} \right \rangle\right)n\left(\gamma,t\right)\right]\\&-\frac{n\left(\gamma,t\right)}{t_{\rm esc}}+Q\left(\gamma,t\right)\,.
    \label{e3}
\end{aligned}
\end{equation}
The left-side term describes the time-dependent evolution of particle distribution $n(\gamma,t)$ in the phase space of energy.

To solve Eq. (\ref{e3}) for both STA and SHA scheme, it is necessary to figure out corresponding Fokker-Planck coefficients $\langle \Delta \gamma/\Delta t \rangle$ and $\langle \Delta \gamma^2/\Delta t \rangle$, which can be obtained through a series of microscopic treatments and approximations \citep{jokipii1990particle,rieger2006microscopic,liu2017particle}. In the following calculations, the accelerated particles will be set as electrons with rest mass $\rm m_e$. The Fokker-Planck relation \citep{rieger2006microscopic} connecting the momentum diffusion coefficient and the momentum variation coefficient can be rewritten in the energy space as:
\begin{equation}
    \langle \frac{\Delta \gamma}{\Delta t} \rangle=\frac{1}{2 \gamma^2}\frac{\partial}{\partial \gamma}\left[\gamma^2\langle \frac{\Delta \gamma^2}{\Delta t} \rangle\right]\,.
    \label{e4}
\end{equation}
For STA, denoting the Alfv$\acute{e}$n velocity of the plasma by $v_{\rm A}$, the diffusion coefficient in the momentum space is proportional to $v_{\rm A}^2p^2/\kappa$ according to the quasi-linear theory \citep{schlickeiser2002conversion}. Here $\kappa={\rm c}^2\tau_{\rm sc}/3$, representing the spatial diffusion coefficient of particles; And $\tau_{sc}$ is the mean scattering timescale, denoting the time between two scattering events: 
\begin{equation}
    \tau_{\rm sc}=\frac{r_{\rm L}^{2-q}}{c \xi \Lambda_{\rm max}^{1-q}}\quad.
    \label{e12}
\end{equation}
With the expression in Eq.~(\ref{e12}), the energy diffusion coefficient takes the form below if Alfv$\acute{e}$n wave is regarded as the only type of Magnetohydrodynamic (MHD) wave that scatter off particles:
\begin{equation}
    \langle \frac{\Delta \gamma^2}{\Delta t} \rangle_{\rm STA}=\frac{\xi \Gamma_{\rm A}^4\beta_{\rm A}^2c}{r_{\rm L}^{2-q}\Lambda_{\rm max}^{q-1}}\gamma^2=A_{\rm STA}\gamma^q\, ,
    \label{e5}
\end{equation}
\begin{equation}
    A_{\rm STA}=\frac{\xi \Gamma_{\rm A}^4 \beta_{\rm A}^2 c}{\Lambda_{\rm max}^{q-1}}\left(\frac{m_{\rm e}c^2}{eB_0}\right)^{q-2}\,.
    \label{e6}
\end{equation}
where $\beta_{\rm A}$ denotes the dimensionless Alfv$\acute{e}$n speed of the jet and $\Gamma_{\rm A}$ represents its bulk Lorentz factor; $\xi=(\delta B/B_0)^2$, reflecting the strength of magnetic turbulence. $r_{\rm L}$ is the Larmor radius of an electron whose energy is $\gamma m_{\rm e} c^2$ and $\Lambda_{\rm max}$ is the longest wavelength of the MHD wave modes. The index $q$ is related to the power-law turbulence spectrum ($W(k)\propto k^{-q}$) of MHD waves \citep{mertsch2011new}, whose typical values include 5/3 (Kolmogorov-type), 3/2 (Kraichnan-type), 1 (Bohm-type) and 2 (`hard-sphere' limit). With Fokker-Planck relation, the energy variation coefficient can be calculated as:
\begin{equation}
    \langle \frac{\Delta \gamma}{\Delta t} \rangle_{\rm STA}=\frac{\left(2+q\right)A_{\rm STA}}{2}\gamma^{q-1}\,.
    \label{e7}
\end{equation}
Then the acceleration timescale can be estimated using $t_{\rm acc,\rm STA}=\gamma/\langle \Delta \gamma/\Delta t \rangle_{\rm STA}$, and we get:
\begin{equation}
    t_{\rm acc,\rm STA}=\frac{2}{\left(2+q\right)A_{\rm STA}}\gamma^{2-q}\,.
    \label{e8}
\end{equation}

For the SHA, the Fokker-Planck coefficients in the energy space are written in following forms \citep{rieger2006microscopic,liu2017particle,rieger2019particle,wang2021particle}:
\begin{equation}
\langle \frac{\Delta \gamma^2}{\Delta t} \rangle_{\rm SHA} = \frac{2}{15}\Gamma_{\rm j}^4\left(\frac{\partial \beta_{\rm j}}{\partial r}\right)^2c^2\tau_{\rm sc}\gamma^2\,,
\label{e9}
\end{equation}
\begin{equation}
\langle \frac{\Delta \gamma}{\Delta t} \rangle_{\rm SHA} = \frac{6-q}{15}\Gamma_{\rm j}^4\left(\frac{\partial \beta_{\rm j}}{\partial r}\right)^2c^2\tau_{\rm sc}\gamma\,,
\label{e10}
\end{equation}
$\beta_{\rm j}$ represents the flow speed in the velocity-shear layer j, and $\Gamma_{\rm j}$ is the corresponding bulk Lorentz factor.

The acceleration timescale for SHA can be obtained with the relation $t_{\rm acc,SHA}$ = $\gamma/ \left \langle \Delta \gamma/\Delta t\right \rangle_{\rm SHA}$. Just as what has been discussed in the previous section, SHA can only happen in regions where velocity gradients exist. This mechanism will not work in the central spine region of the jet. So we average both $\langle \Delta \gamma^2/{\Delta t} \rangle_{\rm SHA}$ and $\langle \Delta \gamma/{\Delta t} \rangle_{\rm SHA}$ over the transverse radius of the jet, in order to simplify the treatment of the particle acceleration within the entire jet (so that we do not need to introduce an additional space dimension in the Fokker-Planck equation). It results in
\begin{equation}
\langle \overline{\frac{\Delta \gamma^2}{\Delta t}} \rangle_{\rm SHA} = \frac{\int_{0}^{R_{\rm jet}}2\pi r\langle \frac{\Delta \gamma^2}{\Delta t} \rangle_{\rm SHA} dr}{\pi R_{\rm jet}^2}\,,
\label{e14}
\end{equation}
and
\begin{equation}
\langle \overline{\frac{\Delta \gamma}{\Delta t}} \rangle_{\rm SHA} = \frac{\int_{0}^{R_{\rm jet}}2\pi r\langle \frac{\Delta \gamma}{\Delta t} \rangle_{\rm SHA} dr}{\pi R_{\rm jet}^2}\,.
\label{e15}
\end{equation}
Substituting Eqs.~(\ref{e1}), (\ref{e9}) into Eqs.~(\ref{e14}) and Eq.~(\ref{e1}), also Eq.~(\ref{e10}) into Eq.~(\ref{e15}), we can get modified Fokker-Planck coefficients, written as:
\begin{equation}
\langle \overline{\frac{\Delta \gamma^2}{\Delta t}} \rangle_{\rm SHA} =
\frac{2}{15}\overline{\Gamma_{\rm j}^4}\left(\frac{\beta_0}{\eta R_{\rm jet}}\right)^2c^2\tau_{\rm sc}\gamma^2 = \bar{A}_{\rm SHA}\gamma^{4 - q}\,,
\label{e16}
\end{equation}

\begin{equation}
\langle \overline{\frac{\Delta \gamma}{\Delta t}} \rangle_{\rm SHA} =\frac{6-q}{15}\overline{\Gamma_{\rm j}^4}\left(\frac{\beta_0}{\eta R_{\rm jet}}\right)^2c^2\tau_{\rm sc}\gamma = \frac{6-q}{2}\bar{A}_{\rm SHA}\gamma^{ 3- q}\,,
\label{e17}
\end{equation}
with
\begin{equation}
     \overline{\Gamma_{\rm j}^4}=\frac{\eta}{2\beta_0}\left[\ln\left(\frac{1+\beta_0}{1-\beta_0}\right)+\frac{2\beta_0}{1-\beta_0^2}\right]-\frac{\eta^2}{1-\beta_0^2}\,.
     \label{e19}
\end{equation}
and
\begin{equation}
    \bar{A}_{\rm SHA}=\frac{2}{15}\overline{\Gamma_{\rm j}^4}(\frac{\beta_0}{\eta R_{\rm jet}})^2\xi^{-1}\Lambda_{\rm max}^{q-1}(\frac{m_{\rm e}}{eB_0})^{2-q}c^{5-2q}\,.
    \label{e18}
\end{equation}
Those averaged coefficients will be used in further calculations.

Particle acceleration may be hindered by energy loss and escape. We assume the energy loss are dominated by $\dot\gamma_{\rm c} \propto -\gamma^2$ type radiative cooling process such as synchrotron radiation and IC radiation in the Thomson regime:
\begin{equation}
    \langle \dot{\gamma_{\rm c}} \rangle=-\frac{\sigma_TB_0^2\gamma^2}{6 \pi m_{\rm e} c}\left(1+X\right)=-A_{\rm c}\gamma^2\,,
    \label{e20}
\end{equation}
where $\sigma_{\rm T}$ is the Thomson cross section of electrons, X=$u_{\rm rad}/u_{\rm B}$ with $u_{\rm B}=B_0^2/(8 \pi)$ and $u_{\rm rad}$ the energy density of the target photon field. Above a few tens of TeV the IC scattering is dominated by the Cosmic Microwave Background (CMB) \citep{khangulyan2014simple}, and hence the value of $u_{\rm rad}$ is set as $4.2 \times 10^{-13}\,\rm erg/cm^3$, corresponding to the local energy density of CMB background. The timescale for the diffusive escape of electrons from the jet can be given by
\begin{equation}
    t_{\rm esc}=\frac{R_{\rm jet}^2}{2\kappa}=A_{\rm esc}\gamma^{q-2}\,,
    \label{e21}
\end{equation}
\begin{equation}
    A_{\rm esc}=\frac{3R_{\rm jet}^2\xi\Lambda_{\rm max}^{1-q}}{2c}\left(\frac{eB_0}{m_{\rm e}c^2}\right)^{2-q}\,.
    \label{e22}
\end{equation}

Note that Eq.~(\ref{e12}) is valid only when the resonant scattering of wave-particle interaction is applied. In this condition, the scattering of a particle is dominated by the wave whose wavelength is comparable to the Larmor radius of the particles ($r_{\rm L} \sim 1/k$, or the first-order resonance for particles of small rigidity) \citep{demidem2020particle, kulsrud2020plasma}. As the energy of particle exceeds $\gamma_{\rm rsn} \equiv {\rm e} B_0 \Lambda_{\rm max}m_{\rm e}^{-1}{\rm c}^{-2}$, its Larmor radius exceed the longest wavelength $\Lambda_{\rm max}$. The wave-particle scattering timescale would then deviate from Eq.~(\ref{e12}), and high-order resonance needs to be considered. It leads to a modification of the expression of the scattering timescale, which can be given by $\tau_{\rm sc} = r_{\rm L}^2(\rm c \xi \Lambda_{\rm max})^{-1}$ (see Appendix \ref{sec:appendixA} for the derivation). The coefficients involving $\tau_{sc}$ (from Eq.~(\ref{e12}) to Eq.~(\ref{e22})) should also be modified correspondingly. 

With all coefficients obtained, we may set to solve the Fokker-Planck equation. Although the exact age of the black hole - jet system in V4641 Sgr remains unknown, considering its B9\uppercase\expandafter{\romannumeral3} companion star \citep{macdonald2014black}, a quasi-continuous jet may exist for a sufficient long time, and thus we consider a quasi-steady state by setting $\partial n(\gamma,t)/\partial t=0$. Substituting Eqs.~(\ref{e16})-(\ref{e21}) into Eq. (\ref{e3}), the equation for SHA scheme is rewritten as: 
\begin{equation}
    \frac{\partial^2 n}{\partial \gamma^2}+C_1\left(\gamma\right)\frac{\partial n}{\partial \gamma}+C_2\left(\gamma\right)n+2Q \bar{A}_{\rm SHA}^{-1}\gamma^{q-4}=0\,,
    \label{e23}
\end{equation}
where
\begin{equation}
C_1(\gamma)=\left(2-q\right)\gamma^{-1}+2A_{\rm c}\bar{A}_{\rm SHA}^{-1}\gamma^{q-2}\,,
\label{e24}
\end{equation}
and
\begin{equation}
\begin{aligned}
    &C_2 \left(\gamma \right)=4A_{\rm c} \bar{A}_{\rm SHA}^{-1}\gamma^{q-3}\\&-\left(2A_{\rm esc}^{-1}\bar{A}_{\rm SHA}^{-1}-2q+6\right)\gamma^{-2}\,.
    \label{e25}
\end{aligned}
\end{equation}
The injection term $Q$ is usually considered as a Dirac-$\delta$ function, such as $Q=\delta(\gamma-\gamma_{\rm eq})$ with $\gamma_{\rm eq}$ defined as the injection energy. Particles are usually injected to the acceleration process from the thermal pool (i.e., $\gamma_{\rm eq}\gtrsim 1$), where STA dominates. SHA starts to take over the STA when $\left\langle \Delta\gamma/\Delta t \right\rangle_{\rm STA} = \left\langle \Delta\gamma/\Delta t\right\rangle_{\rm SHA}$, or at $\gamma=\gamma_{\rm eq}$ with
\begin{equation}
    \gamma_{\rm eq}=\left[\frac{\left(2+q\right)A_{\rm STA}}{\left(6-q\right)\bar{A}_{\rm SHA}}\right]^{\left(\frac{1}{4-2q}\right)},\,\left(q \neq 2\right)\,.
    \label{e26}
\end{equation}
Therefore, we simply consider $Q=0$ in Eq.~(\ref{e23}) since we focus on the acceleration of TeV-PeV particles.

At $\gamma<\gamma_{\rm eq}$, the STA dominates and the accelerated spectrum can be given by \citep{liu2017particle}: 
\begin{equation}
    N_{\rm 1}\left(\gamma\right)=-\frac{2C\gamma}{\left(1+q\right)A_{\rm STA}\gamma^q}\propto\gamma^{1-q}\,,
    \label{e28}
\end{equation}
which is a single power-law spectrum.

\citealt{wang2021particle} analytically solve Eq.~(\ref{e23}) with $Q=0$ for $1<q\le2$: Substituting trial solutions with the form of $n(\gamma)\propto\gamma^sj(\gamma)$ and introducing a new variable $z=(6-q)/(1-q)(\gamma/\gamma_{\rm max})^{q-1}$ ($\gamma_{\rm max}$ equals to the $\gamma$ value where $t_{\rm acc,\rm SHA}$ = $t_{\rm c}$), $j(z)$ can be expressed in a Kummer-form differential equation, whose general solutions are combinations of confluent hyper-geometric functions, denoted by $_{1}F_1$ \citep{abramowitz1948handbook}. The final solution for $n(\gamma)$ is:
\begin{equation}
    N_{\rm 2}(\gamma)=K_1\gamma^{s_+}{_1F}_1\left(a_+,b_+;z\right)+K_2\gamma^{s_-}{_1F}_1\left(a_-,b_-;z\right)\,,
    \label{e27}
\end{equation}
where $a_\pm=(2+s_\pm)/(q-1)$, $b_\pm=2s_\pm/(q-1)$, $s_\pm=(q-1)/2\pm \sqrt{(5-q)^2/4+w}$, $w=(6-q)t_{\rm acc,\rm SHA}/t_{\rm esc}$, and both $K_1$ and $K_2$ are normalization factors. In our spectral model, $w=10\eta^2/(\beta_0^2\overline{\Gamma_{\rm j}^4})$. Eq.~(\ref{e27}) is valid to describe the accelerated electron spectrum for $\gamma_{\rm eq} < \gamma < \gamma_{\rm rsn}$, with the low-order resonance being satisfied. Note that $s_+$ is always larger than 0, so we have $K_1=0$ based on the boundary condition $N_2(\gamma\to \infty)\to 0$.

At $\gamma>\gamma_{\rm rsn}$, the particle spectrum is subject to change due to the modified scattering rate. The solution for the Fokker-Planck equation of SHA at this energy range is found to be in a power-law form:
\begin{equation}
    N_{\rm 3}(\gamma) = D_1 \gamma^{p_{\rm +}} + D_2 \gamma^{p_{\rm -}}\,,
    \label{esol}
\end{equation}
where $p_{\rm \pm} = -1/2 \pm \sqrt{25/4 + w}$. The complete derivation is shown in Appendix \ref{sec:appendixB}. Similarly, $p_+>0$ is always established so we have $D_1=0$ based on the boundary condition $N_3(\gamma\to \infty)\to 0$.

Combining Eq. (\ref{e28}) Eq. (\ref{e27}) and Eq. (\ref{esol}), the model for electron spectrum considering both stochastic and shear acceleration can be summarized as the piecewise function shown below:
\begin{equation}
N(\gamma)=
\left \{
\begin{array}{l@{\quad}l}
\begin{aligned}
&K_0\gamma^{1-q}&  \gamma<\gamma_{\rm eq}\,, \\
&K_2\gamma^{s_-}{_1F}_1\left(a_-,b_-;z\right) & \gamma_{\rm eq}\le \gamma<\gamma_{\rm rsn}\,,\\
&D_2\gamma^{p_{\rm -}} &\gamma_{\rm rsn}\leq\gamma\,.
\label{e29}
\end{aligned}
\end{array}
\right.
\end{equation}
To maintain the continuity of the spectrum, we have $N_1(\gamma_{\rm eq}) = N_2(\gamma_{\rm eq})$ and $N_2(\gamma_{\rm rsn}) = N_3(\gamma_{\rm rsn})$ so that the normalization factors could be obtained. Given the boundary condition, the normalization coefficients are replaced by a single free parameter indicating the total number of accelerated electrons: $N_{\rm tot}=\int_{\gamma_{\rm min}}^{+\infty}N(\gamma)d\gamma$, which efficiently simplifies the parameter space. Here $\gamma_{\rm min} \sim 2$, corresponding to MeV electrons.  

\subsection{Parameter constraints}
\label{paracons}
{The cooling term and the escaping term in the Fokker-Planck equation serve as natural constraints on the parameters, for the spectrum would be quite steep if the two terms become dominated. In order to preliminarily narrow down the parameter space, the influences of cooling and diffusive escape should be discussed. 

By equating the acceleration rate Eq.~(\ref{e17}) and the cooling rate Eq.~(\ref{e20}) , the maximum energy $\gamma_{\rm max}$ is achieved when $t_{\rm acc,\rm SHA}=t_{\rm c}$, written as:}
\begin{equation}
    \gamma_{\rm max}=\left[\frac{\left(6-q\right)\bar{A}_{\rm SHA}}{2A_{\rm c}}\right]^{\frac{1}{q-1}}\,.
    \label{e32}
\end{equation}
If $\gamma>\gamma_{\rm max}$, the spectrum would present a cutoff due to the effects of cooling. As to the diffusive escape, it becomes more efficient at higher energy as long as $1 < q \le 2$. For the SHA, the dependence of the acceleration timescale and the escape timescale on particle energy is the same. If the escape timescale is shorter than the acceleration timescale, the accelerated spectrum would be significantly softened.  In order to maintain a hard electron spectrum, the escape timescale should not be much shorter than that for the SHA, i.e. $t_{\rm acc,\rm SHA}\lesssim t_{\rm esc}$ 
    \begin{equation}
        \eta\lesssim [\beta_0\ln\left(\frac{1+\beta_0}{1-\beta_0}\right)+\frac{2\beta_0^2}{1-\beta_0^2}]\left(\frac{20}{6-q}+\frac{2\beta_0^2}{1-\beta_0^2}\right)^{-1}\quad.
        \label{e33}
    \end{equation}
{Eq. (\ref{e33}) tells that the velocity-shear region should be small enough to achieve a sufficiently high acceleration efficiency.

Concerning that when $q$ = 5/3, the STA spectrum becomes too hard to fit to the observed spectrum, we speculate that in this case the spectrum above a TeV would not be dominated by STA. This requires that $\gamma_{\rm eq} m_{\rm e} c^2$ should not significantly exceed 1 TeV to ensure the domination of SHA above 1 TeV:
\begin{equation}
    A_{\rm STA}\lesssim \frac{(6-q)\bar{A}_{\rm SHA}}{(2 + q)}\gamma^{(4-2q)}\,,
    \label{1TeV}
\end{equation}
for $\gamma\sim 1{\rm TeV}/(m_ec^2)$. Once $n_{\rm p}$ becomes too small, STA would turn out to be more efficient due to a larger Alfv$\acute{e}$n speed ($\beta_A = B_0/\sqrt{4 \pi n_{\rm p}m_{\rm p}}$) and in this case Eq. (\ref{1TeV}) could no longer be satisfied.

Aside from these constraints, the observations of V4641~Sgr pose additional restrictions on the parameters, which are listed as follows:
\begin{enumerate}
    \item \textbf{The maximum luminosity}: According to historical X-ray observations, the most luminous flare from the microquasar in 1999 reached a luminosity of $\sim (3-4) \times 10^{39}\, \rm erg/s$ over 1-10\,keV \citep{revnivtsev2002super}, which is about three times the Eddington luminosity of a 10$M_{\rm sun}$ black hole ($L_{\rm Edd}$). As a conservative estimate, the long-term average kinetic luminosity of the quasi-continuous jet, defined as the total kinetic energy of particles flowing through the jet's cross section per second, should not significantly exceed the Eddington luminosity:
    \begin{equation}
        L_{\rm kin}=\pi (1-\eta)^2R_{\rm jet}^2n_{\rm p} \Gamma \left(\Gamma-1\right)m_{\rm p} c^2\left(\beta_0c\right)\le L_{\rm Edd}\ ,
        \label{e30}
    \end{equation}
   where $\Gamma=1/\sqrt{1-\beta_0^2}$ is the bulk Lorentz factor of the jet's spine (here we neglect the kinetic luminosity of the velocity-shear region); $n_{\rm p}$ represents the number density of protons in the jet. 
   Assuming $L_{\rm Edd}=1.3\times10^{39}\,$erg/s, Eq.~(\ref{e30}) gives the approximation for the maximum protons' number density of the jet: 
    \begin{equation}
        n_{\rm p} \lesssim 4 \times10^{-8}\left[\Gamma(\Gamma-1)\right]^{-1}\beta_0^{-1}(1-\eta)^{-2}\left(\frac{R_{\rm jet}}{5\,\rm pc}\right)^{-2}\,\rm cm^{-3}.
        \label{e31}
    \end{equation}
    
    \item \textbf{Confinement of particles}: During acceleration processes, the electrons should be confined in the acceleration region. That is to say, their gyro-radius $r_{\rm L}$ should be smaller than the characteristic length of the accelerator $R_{\rm jet}$, which is roughly equivalent to the Hillas criterion with $\beta_0\lesssim1$ \citep{hillas1984origin}. It therefore provides a constraint on the maximum particle energy by 
    \begin{equation}
        \gamma_{\rm Hillas} = \frac{eB_0R_{\rm jet}}{m_{\rm e} c^2}\,.
    \end{equation}
    If $\gamma>\gamma_{\rm Hillas}$, the electrons will quickly escape the jet, leading to a cutoff in the spectrum around $\gamma_{\rm Hillas}$. Moreover, if the mean free path (MFPs, $\lambda$) of the electrons is comparable to or longer than the radius of the jet ($\lambda = c\tau_{\rm sc}<R_{\rm jet}$), particles can escape the jet without being scattered and cannot be described with the Fokker-Planck equation, leading to a cutoff in the accelerated spectrum around the energy $\gamma_{\rm MFP}$ which is defined by $c\tau_{\rm sc}=R_{\rm jet}$ or
    \begin{equation}
        \gamma_{\rm MFP} = \frac{eB_0}{m_{\rm e} c^2}\left( \xi \Lambda_{\rm max}^{1-q}R_{\rm jet}\right)^{\frac{1}{2-q}}\,,
        \label{e35}
    \end{equation}
    where $\Lambda_{\rm max}=\eta R_{\rm jet}$ is employed in the calculation, i.e., the longest wavelength of turbulence is equal to the transverse size of the velocity-shear region. This is a conservative assumption because the $\Lambda_{\rm max}$ can be in principle comparable to the entire size of the jet and the constraint on the maximum particle energy may be relaxed. This constraint can be also expressed by $\xi^{-1}(r_{\rm L}/\Lambda_{\rm max})^{1-q}r_{\rm L}<R_{\rm jet}$. The constraint on the maximum energy is generally stronger than the Hillas criterion for relativistic flow ($\beta_0\lesssim 1$), unless in the Bohm limit ($\xi=1$ and $q=1$).
    
\end{enumerate}

With those constraints, we can roughly figure out the range of some parameters. For a leptonic interpretation of the LHAASO observation, electrons must be accelerated to at least 0.8\,PeV. Given the above equations being satisfied, constraints on the magnetic field and $\eta$ could be obtained, which are shown in Figure~\ref{fig:confine}. According to calculations, $B_{\rm 0}$ could vary within a wide range, from less than 0.1 $\rm \mu$G to around 100 $\rm \mu$G when $R_{\rm jet}$ = 5 pc; Eq. (\ref{e31}) indicates that the jet can not be ultra-relativistic, or the upper limit of the density might be smaller than its lower limit. This leads to another conclusion that the size of the velocity-shear region is limited to a relatively small range. 

\section{Results} \label{sec:NonThermal}

\subsection{Fitting to the gamma-ray data} \label{subsec:UHE}
We then fit the observational data with derived spectral model, taking into account constraints obtained in section \ref{paracons}. Our spectrum model contains four main free parameters: $B_0$, $\eta$, $\beta_0$, and $N_{\rm tot}$. For other parameters, $\xi$ is set to 1, denoting that $\delta B_0 \sim B_0$. It is a reasonable trial estimation to replace the mean field with the turbulent field if the latter dominates over the former, indicating that the induced turbulent field has achieved a ``saturation'' level \citep{amato2009kinetic, WangJS2023}. MHD simulations on the formation of velocity-shear sheath in jets also prove that the strength of the turbulence can reach this level during the evolution \citep{WangJS2023}. The index of the turbulence spectrum ($q$) is set to 5/3, corresponding to the Kolmogorov-type turbulence. $\Lambda_{\rm max}$ is given by $\Lambda_{\rm max}=\eta R_{\rm jet}$. In our model, shear-accelerated electrons are responsible for observed TeV - PeV photons via IC scattering. \texttt{naima}, a python package which can compute the non-thermal radiation from given electron populations \citep{zabalza2015naima} is used to calculate the IC emission produced by the modeled electron distribution. Implements precise analytical solutions for the IC scattering of black-body seed photons \citep{khangulyan2014simple}, which is computationally cheap and can achieve high precision under both the Thomson and KN limits. Here three types of seed photon field are selected to calculate the radiation, including CMB, Far-Infrared (FIR) and Near-Infrared (NIR).

The inclination angle of the jet is unclear. However, results from the ellipsoidal orbital fittings with the ELC code \citep{macdonald2014black} show that the inclination angle \textit{i} of the source, which is defined as the angle between the direction perpendicular to the binary orbital plane and the observer's line of sight, is ${72.3\pm 4.1}^{\circ}$. In a standard binary system, the accretion disk is parallel with the orbital plane due to the conservation of momentum and the jet is nearly perpendicular to the disk \citep{blandford1977electromagnetic}, so we set the jet angle as $72.3^{\circ}$ when computing the relativistic beaming effect to match the recent UHE observations.

For a source distance of $d=6.2\,$kpc, the attenuation of high-energy gamma-ray photons need be taken into consideration, because the typical attenuation length of PeV photons absorbed by CMB is $\lesssim 8\,$kpc \citep{Coppi1997,dermer2009high}, comparable to the source distance. So we compute the optical depth $\tau_{\gamma \gamma}$, based on which we correct the results of \texttt{naima} to obtain the observed flux. 

Markov Chain Monte Carlo (MCMC) is a powerful algorithm to find the best-fit parameters for a model basing on Bayesian methods. We use the Python package \texttt{emcee} to carry out the fittings. The observational data mainly comes from two experiments: LHAASO \citep{lhaaso2024ultrahigh} and HAWC \citep{alfaro2024ultra}. To explore the parameter space more efficiently, all parameters, except $\eta$ and $\beta_0$, are redefined in their logarithmic forms. The autocorrelation time is calculated to ensure convergence of the results, and constraints in section \ref{paracons} are used to define the prior function. Data collected from H.E.S.S. observatory, ranging from 1 to 40 TeV is extracted from \citealt{neronov2025multimessenger} to further constrain the model. All parameters and their best-fit values are shown in Table \ref{tab:paras1}. The UHE spectrum can be well reproduced with the IC radiation of accelerated electrons under all $R_{\rm jet}$ considered, as shown in Figure \ref{fig:IC}. 

Within the framework of SHA, the obtained magnetic field depends on the assumed radius of the jet. A larger $B_0$ is required for a smaller $R_{\rm jet}$, due to the requirement of confinement as discussed in section~\ref{paracons}. On the other hand, the velocity of the jet spine keeps more or less the same ($\sim 0.6\rm c-0.7 \rm c$) for different $R_{\rm jet}$, which is lower than the fitting results from \citealt{hjellming2000light} ($\sim$ 0.85$\rm c$). This discrepancy might arise from that the propagated jet has been decelerated. Based on these parameters, the baryon density of the jet $n_{\rm p}$ may range from $2.5\times 10^{-11}~\rm cm^{-3}$ to $2.6\times10^{-7}~\rm cm^{-3}$ for $R_{\rm jet}=5$\,pc (the largest $R_{\rm jet}$ employed), and from $3.6\times 10^{-9}\,\rm cm^{-3}$ to $3.0\times 10^{-5}\,\rm cm^{-3}$ for $R_{\rm jet}=0.5\,$pc (the smallest $R_{\rm jet}$ employed), according to the constraints from Eqs.~(\ref{1TeV}) and (\ref{e30}). Note that $n_{\rm p}$ has little impact on the obtained electron spectrum as long as it is within the reasonable range. The wide available range of $n_p$ shows that the model parameters are self-consistent. Such low proton densities in the jet also indicates that the hadronic emission in our model could be quite weak though a small portion of the protons might be accelerated. The number density of target protons for pp collisions could be assumed as $\sim$ $10^{-8}$\,$\rm cm^{-3}$, which is a typical density derived from the model. Then the timescale for $pp$ collision is estimated to be $t_{pp}\sim$ $10^{23}$ s, which is much longer than the kinetic timescale for the system. Even if relativistic protons are produced with a luminosity of $L_p=10^{37}$erg/s over the past 100\,kyr, which leads to a total relativistic proton energy to be $W_p=3\times 10^{49}\,\rm erg$, the luminosity for hadronic component is estimated to be $\sim W_p/t_{pp}\sim 3\times 10^{26}\rm \, erg/s$, which is nearly 7 orders of magnitude smaller than the observed gamma-ray luminosity. As a result, the expected gamma-ray flux of hadronic origin may be neglected.

In section \ref{sec: jetmodel_spec}, we have obtained all timescales from Eq. (\ref{e12}), (\ref{e7}), (\ref{e17}), (\ref{e20}) and Eq. (\ref{e21}). With the best-fit parameters for the model, all timescales are shown in Figure \ref{fig:times}. As we expected, STA dominates the acceleration at low energies ($\gamma<\gamma_{\rm eq}$), providing `seed particles', and SHA becomes more important at $\gamma>\gamma_{\rm eq}$. Acceleration of particles around $\gamma\sim\gamma_{\rm eq}$ takes the longest time, which is about a few times $10^{10}$\,s for all four cases with varying $R_{\rm jet}$ values. It implies that the system must exist for at least a few thousand years, or otherwise we cannot apply the quasi-steady state solution. 
The jet active time can be constrained by the morphology of the system, which depends on the jet-ambient medium interaction \citep[e.g, see Appendix A of][for a short summary of interaction models]{Wang:ApJL:2025}.
Here we adopt a simple scenario assuming a continuous and cylindrical cold jet. 
The jet head velocity is determined by $\beta_h=\beta_0\sqrt{\zeta}/(1+\sqrt{\zeta})$ and $\zeta=\Gamma^2n_{p}/n_{\rm ext}$, where $n_{\rm ext}$ is the ISM density in units of cm$^{-3}$ \citep{Marti:ApJ:1997}. 
For a jet length of $D=100~$pc, the required active time is $T= D/(\beta_h c)\approx2 n_{\rm ext}^{1/2}R_{\rm jet,pc} L_{\rm kin,37}^{-1/2}$~Myr assuming $\zeta<1$, where $R_{\rm jet}= 1 R_{\rm jet,pc}$~pc and $ L_{\rm kin}=L_{\rm kin,37}10^{37}~{\rm erg/s}$.
It can be seen that for a jet with pc-scale radius, an active time of $200$~kyr is required for a low-density region of $n_{\rm ext}=0.01~{\rm cm}^{-3}$ and a power of $10^{37}~{\rm erg/s}$.
For such a duration of time, the jet is able to propagate to $\sim 100\,$pc away from the BH with a mildly-relativistic velocity, although the speed of the jet head could have been decelerated to non-relativistic.  We also find that $t_{\rm acc,\rm SHA} \lesssim t_{\rm esc}$ in all four cases. This is not a coincidence. From Fig.~\ref{fig:IC} we see that the gamma-ray photon index from 1\,TeV to a few times 100\,TeV (before the softening/cutoff occurs) is approximately $\lesssim -1.5$, so we need approximately $N(\gamma)\propto \gamma^{-2}$ to reproduce the observed spectrum, or $s_-=(q-1)/2-\sqrt{(5-q)^2/4+w}=-2$. This gives $t_{\rm acc, SHA}/t_{\rm esc}=w/(6-q)=4(q-1)/(6-q)$. For $1<q\leq 2$, we generally have $t_{\rm acc, SHA}/t_{\rm esc}\lesssim 1$. $t_{\rm acc, SHA}/t_{\rm esc}\lesssim 1$ implies a relatively efficient escape of accelerated electrons. These electrons may produce a halo-like UHE structure in the surrounding ISM via IC radiation, making the source more extended than the transverse size of the jet.


\subsection{X-ray synchrotron emission} \label{subsec:Xray}
In the emission region, accelerated electrons will also generate synchrotron radiation when traversing the magnetic field. For TeV - PeV electrons in a magnetic field at the micro-gauss level, the radiation should peak at the X-ray band. Within the SHA scenario, we can predict the X-ray synchrotron emission from the electrons which are responsible for the detected UHE emission. The \texttt{naima} package is also used to calculate the synchrotron emission, based on the emissivity function in random magnetic fields \citep{aharonian2010angular}. The results are shown in Figure \ref{fig:SYN}. With the best-fit parameters, the unabsorbed total X-ray flux in $2 - 10$\,keV varies from $10^{-13}$ to $10^{-12}\rm \,erg\,cm^{-2}\,s^{-1}$ depending on $R_{\rm jet}$ and $B_0$.
Since the UHE gamma-ray and the non-thermal X-ray emission have the same origin in the leptonic model, we would expect an elongated morphology of the X-ray emission as well. 

Recent observation of XRISM has revealed an extended X-ray source around V4641 Sgr, which may constrain the model. The observation focuses on the central part of the microquasar, and the spatial extension is much smaller than the UHE structure detected by HAWC and LHAASO. They argued that if the X-ray emission is of non-thermal origin, its original electron population should be different from that responsible for the gamma-ray emission. Therefore, the surface brightness of the XRISM source, which is around $S_{\rm X}=(7-10) \times 10^{-15} \rm erg\,s^{-1}\,cm^{-2}\,arcmin^{-2}$ in 2-10\,keV \citep{suzuki2025detection}, sets up an upper limit of the synchrotron emission of accelerated electrons predicted in our model. We calculate the $2 - 10$\,keV flux from the whole jet, and compare it with the X-ray flux upper limit set by XRISM observation $S_{\rm X}\Omega$. Here, $\Omega\approx 2R_{\rm jet}\times 100\,{\rm pc}/d^2 = 307\,(R_{\rm jet}/5\,\rm pc) \,\rm arcmin^2$ is the solid angle of the jet subtended in the sky, with $100\,$pc being the projected length of the jet and $d=6.2\,$kpc being the distance of the microquasar. The X-ray flux upper limit for different $R_{\rm jet}$ are shown in Figure \ref{fig:SYN}. We see that for $R_{\rm jet} = 5$\,pc and 3\,pc, the predicted X-ray flux from the model is lower than the upper limit. When $R_{\rm jet}$ decreases to 1\,pc, the predicted X-ray flux overshoots the upper limit even considering the lower bound of the parameter uncertainties. So we may conclude that $R_{\rm jet}>1\,$pc is favored for this model. Note that since XRISM observations actually do not cover the full UHE emission region, the constraint may be relaxed if future observations detect X-ray emission from UHE gamma-ray-emitting region outside its field of view.

\section{Discussion} \label{sec:Discus}
\subsection{Dependence on the turbulence spectrum $q$}
Figures~\ref{fig:IC} and \ref{fig:SYN} present the results by assuming a Kolmogorov-type turbulence with $q=5/3$. Actually, for other types of turbulence spectrum, the results will not change significantly unless for $q\to 1$. In Figure~\ref{fig:VariableQ}, we show the fitting to the gamma-ray spectrum and the predicted X-ray flux with $q=3/2$ and $q=2$, assuming $R_{\rm jet}=3\,$pc. The main parameters do not vary much from those with $q=5/3$, as shown in Table.~\ref{tab:paras2}, except for $N_{\rm tot}$ with $q=2$. The reason is that the timescales of STA and SHA, as well as the escape timescale, all become energy-independent for $q=2$. In this case, when the SHA dominates both low-energy and high-energy end, the spectrum is softened by the escape 
(see discussion in Section~\ref{subsec:UHE}) and the accelerated spectrum becomes softer than $\gamma^{-1}$. The total electron number then significantly increase by integrating the spectrum down to $\gamma_{\rm min}=2$. Nevertheless, this corresponds to a total electron kinetic luminosity of $\sim 10^{36-37}\,\rm erg/s$, which is still reasonable.  On the other hand, for $q\to 1$ the energy of the turbulence cascaded down to the gyro-scale of electrons of relevant energies is much higher than those with larger $q$. As a consequence, the mean free path of electrons becomes smaller and hence the
acceleration timescale of SHA will become longer (i.e., the SHA becomes less efficient). We find a higher jet velocity of $\beta_0>0.95$ is needed with $q=1$. The relativistic effect becomes more important and the analytical solution derived in Section~\ref{sec: jetmodel_spec} may need modification.

\subsection{Can STA work solely?} \label{sec:PureStc}
STA is also a `distributed' acceleration, which may be helpful to explain the elongated morphology of UHE emission around V4641~Sgr in the leptonic scenario. In section~\ref{paracons}, we discussed that the accelerated particle spectrum of STA is too hard to explain the measured spectrum of V4641~Sgr. However, with $q=2$, the energy dependence of the acceleration timescale is the same with that of the escape timescale. In this case, when two timescales become comparable, the accelerated spectrum can be softened (similar to the SHA case) and the observed spectrum may be reproduced by pure STA with fine-tuning parameters. The main issue is the maximum particle energy achievable by STA. Indeed, strong turbulence and high Alfv{\'e}n speed need be present in the environment for efficient STA, while observations of XRISM constrain the magnetic field strength to be lower than $\sim 1\,\mu$G in the jet. Therefore, a very low density would be needed to get a high Alfv{\'e}n speed. From Figure~\ref{fig:SYN}, we see that there is still some room for a higher magnetic field in the cases of $R_{\rm jet}=5\,$pc and $3\,$pc, which can be as high as $2\,\mu$G from the perspective of X-ray upper limit. On the other hand, a smaller $\Lambda_{\rm max}$ would lead to more efficient STA as long as the 1st-order resonance condition for 0.8\,PeV electrons can be satisfied, i.e., $\gamma_{\rm rsn} \equiv {\rm e} B_0 \Lambda_{\rm max}m_{\rm e}^{-1}c^{-2}\geq 0.8\,{\rm PeV}/\left(m_{\rm e}c^2\right)$ (otherwise, the STA becomes less efficient with high-order resonance). We take an optimistic setup for STA, i.e., $\xi = 1$, $B_0=2\,\mu$G, and $\Lambda_{\rm max}=0.8\,{\rm PeV}/\left(eB_0\right)$. Given these conditions, we require the acceleration timescale of STA to be shorter than the synchrotron cooling timescale for 0.8\,PeV electrons, and we have
\begin{equation}
    \Gamma_{\rm A}^4\beta_{\rm A}^2\geq A_{\rm c}\Lambda_{\rm max}\gamma\left(c\xi\right)^{-1}\
\end{equation}
for $\gamma=0.8\,{\rm PeV}/\left(m_ec^2\right)$. We find that in this case, $\beta_{\rm A}$ need be larger than 0.09, which is translated to $n_{\rm p}<2.4\times 10^{-8}\,\rm cm^{-3}$. A very low baryon density of the jet is needed.

\subsection{The jet-orbit misalignment}
In our calculations, the jet inclination angle $\theta_{\rm jet}$ is set as the same value with $i$, based on the theoretical framework where relativistic jets are launched along black hole spin axes \citep{blandford1977electromagnetic}, with the spin axis orientation assumed to be perpendicular to the accretion disk plane. Notably, during its 1999 flaring state, V4641 Sgr exhibited a transient radio jet with highly relativistic bulk motions at the small scale, implying that the jet could be nearly aligned with the observer’s line of sight at launch \citep{orosz2001black}. Subsequent Chandra observations further reveal that broad X-ray emission lines persists over four orders of magnitude in the X-ray luminosity, consistent with Doppler broadening from a high-speed jet viewed at low inclination \citep{gallo2013v4641}. These scenarios introduce a conflict that the inferred low jet inclination angle ($\theta_{\rm jet}<10^{\circ}$ ) is significantly misaligned with the orbital plane’s normal, posing a puzzling contradiction. While results from \citealt{hjellming2000light} indicate that the jet's inclination angle could be around $63^{\circ}$, which is more consistent with the one adopted in our model.

In fact, the jet-orbit misalignment, or spin-orbit misalignment has been found in several binary systems, such as Cygnus X-1 \citep{zdziarski2023evidence} and MAXI J1820+070 \citep{poutanen2022black}. Among all of them, V4641 Sgr remains special for its large misalignment angle ($>$ $50^{\circ}$). Models such as precessing jets \citep{gallo2013v4641} and natal kick evolution \citep{salvesen2020origin} have been proposed to explain the spin-orbit misalignment of the source; However, the former remains to be unproven and the latter struggles to account for such a large misalignment angle. Simulations from population synthesis (PS) models also tell that only a minor part of the XRB population ($<5\%$) could reach $>20^{\circ}$ misalignment angles \citep{fragos2010black}. So far, the exact origin of the misalignment is still under debate. Nevertheless, considering the mild-relativistic jet with $\Gamma \gtrsim 1$ and weak beaming effects, the actual inclination angle of the jet would not bring significant impacts on our results.

\section{Conclusions}
\label{sec:Conclusions}
In this paper, we investigated the possibility of interpreting the UHE gamma-ray emission of V4641~Sgr with a leptonic model. We speculated that the elongated UHE morphology of V4641~Sgr arises from electrons accelerated along the jets of the microquasar. Such a kind of `distributed' acceleration may be achieved with the shear acceleration. It has been suggested that jets with velocity-shear flows have the potential to accelerate electrons to $\sim$ PeV level with this mechanism given favorable parameters. We apply the mechanism to V4641~Sgr. The electron spectrum is obtained by analytically solving the Fokker-Planck equation, with taking into account the pre-acceleration of particles by the stochastic acceleration, cooling and diffusive escape of particles, and high-order resonance scattering between wave and particles. We find that the observed gamma-ray spectrum can be well reproduced with the IC radiation of accelerated electrons generated by the spectral model. Using the MCMC algorithm, the obtained magnetic field inside the jet under four chosen values of $R_{\rm jet}$ are convergently constrained at $\mu$G level and the bulk jet speed is around $0.6\rm c-0.7\rm c$. We also calculated the X-ray synchrotron emission from the same electron population, whose total flux in $2-10$\,keV could range from $10^{-13}$ to $10^{-12}\rm \,erg\,cm^{-2}\,s^{-1}$. Recent observations of XRISM \citep{suzuki2025detection} on V4641~Sgr may favor $R_{\rm jet} > 1$\,pc. 

Future observations in X-ray band and TeV-PeV gamma-ray band are crucial to test the model.
Compared with the hadronic models of escaping protons \citep{ohira2024very, neronov2025multimessenger}, our leptonic model provides a relatively strong constraint on the magnetic field of the jet and future X-ray observations of full coverage of the UHE source may give a stronger constraint on the model. Also, our model would predict a truly elongated morphology of the UHE gamma-ray source, whereas the jet-termination scenario or two-lobe scenario \citep{alfaro2024ultra} would predict two separate sources. The morphology can be hopefully measured at a better precision with future observations of imaging air Cherenkov telescopes such as LACT \citep{LACT2024}, CTA \citep{CTA2011}, and ASTRI \citep{ASTRI-MINI22}. 

\appendix

\section{The high-order resonance scenario}
\label{sec:appendixA}
In the cosmic-ray transport theories, the spatial diffusion coefficient $\kappa$ is linked with the pitch-angle coefficient $D_{\mu\mu}$, where $\mu$ represents the cosine angle between the particles and the magnetic field lines \citep{jaekel1992fokker}:
\begin{equation}
    \kappa \propto \int_{0}^{1}\frac{\left(1-\mu^2\right)^2}{D_{\mu\mu}}d\mu\,.
    \label{e40}
\end{equation}
$D_{\mu\mu}$ for the idealized case can be obtained with quasi-linear theory, written as Eq. (\ref{e41}) \citep{demidem2020particle} if the MHD wave is dominated by Alfv$\acute{e}$n modes:
\begin{equation}
\begin{aligned}
    D_{\mu\mu}^A =&\frac{\Omega^2 \pi^2\left(1-\mu^2\right)}{2B_0^2}\sum_{\pm}\sum_{n = -\infty}^{n = \infty}\int\int dk_{\bot}dk_{\parallel}k_{\bot}\\ 
    & \times \delta \left(k\mu_k\mu-\omega_{\rm A}+n\Omega\right)\left[J_{n+1}\left(z_{\bot}\right)+J_{n-1}\left(z_{\bot}\right)\right]^2k_{\parallel}^{-2}\\
    &\times \left(k_\parallel-\mu\omega_{\rm A}\right)^2S_k^{\rm A}\,,\\
    &z_{\bot}=\frac{\left(k\sqrt{1-\mu_k^2}\right)\sqrt{1-\mu^2}}{\Omega}\,,\\
    &\Omega=\frac{c}{r_{\rm L}},\quad \omega_{\rm A}=\pm \beta_{\rm A} k _{\parallel},\quad k_\parallel=k\mu_k\,,
    \label{e41}
\end{aligned}
\end{equation}
where $k$ represents the wave number vectors and $\mu_k$ is the cosine angle between the wave vectors and the magnetic field lines. $k_{\parallel}$ and $k_{\bot}$ are parallel and vertical components of $k$ with respect to the $B_0$ direction, respectively. $S_k^{\rm A}$ is defined as the power spectrum of the turbulence, and the resonance function is in the idealized, undamped form. The longest wavelength $\Lambda_{\rm max}$ can be transformed to the minimum wave number $k_{\rm min}$. Noting that $k>k_{\rm min}$ and the resonance condition requires that $k\mu_k\mu-\omega_{\rm A}+n\Omega = 0$, we have $ \left | n \right |$ $>$ $r_{\rm L}k_{\rm min}$. It is evident that when $r_{\rm L}<\Lambda_{\rm max}$ ($r_Lk_{\rm min}<1$), the low-order resonant terms will take over, making it easier for particle-wave scatterings to happen. While if $r_{\rm L}>\Lambda_{\rm max}$ ($r_{\rm L}k_{\rm min}>1$), only large n values are allowed. As a result, the wave-particle resonance becomes weaker. 

After applying the Goldreich-Sridhar power spectrum (see Eq. (12) from \citealt{ demidem2020particle}) and summing up all the numerical integrations, we find that $D_{\mu\mu} \propto (r_{\rm L}k_{\rm min})^{-2}$, which is coherent with the results obtained in \citealt{demidem2020particle}. It indicates that $\kappa \propto r_{\rm L}^2 $ when $r_{\rm L}>\Lambda_{\rm max}$. Assuming $r_{\rm L}(\gamma_{\rm rsn})$ = $\Lambda_{\rm max}$, for more energized electrons ($\gamma>\gamma_{\rm rsn}$), their MFPs could be rewritten as: 
\begin{equation}
\lambda =c\tau_{\rm sc} = \left(\xi\Lambda_{\rm max}\right)^{-1}r_{\rm L}^2\,,
\label{modified_MFP}
\end{equation}
causing a higher rate for both SHA and escape, along with an earlier cutoff due to the confinement of MFP.

\section{Derivation of the modified spectrum}
\label{sec:appendixB}
When $r_{\rm L} > \Lambda_{\rm \max}$, substituting Eq. (\ref{modified_MFP}) into the expressions of the coefficients, Eq. (\ref{e16}), Eq. (\ref{e17}) and Eq. (\ref{e21}) becomes:
\begin{equation}
\begin{aligned}
    & \left \langle \frac{\Delta \gamma^2}{\Delta t} \right \rangle_{\rm SHA} = \bar{A}_{\rm SHA} \gamma^4 \,,\\
    &\left \langle \frac{\Delta \gamma}{\Delta t}\right \rangle_{\rm SHA} = 3\bar{A}_{\rm SHA}\gamma^3 \,,\\
    & t_{\rm esc} = A_{\rm esc}\gamma^{-2}\,.\\
\end{aligned}
\end{equation}
According to Figure \ref{fig:times}, the cooling timescale becomes much longer than acceleration timescale in this case, thus the cooling term can be neglected. With $\left \langle \dot{\gamma_{\rm c}}\right \rangle$ = 0, Eq. (\ref{e3}) turns to the form:
\begin{equation}
    \gamma^2 \frac{\partial^2 n}{\partial \gamma^2} + 2\gamma\frac{\partial n}{\partial \gamma} -2 \left(3 + \frac{2}{A_{\rm esc}\bar{A}_{\rm SHA}}\right)n = 0\,,
\end{equation}
which is a typical Cauchy-Euler equation. Substituting the trial solution $n\propto$ $\gamma^r$, we have the relation:
\begin{equation}
    \left[r^2 + r -2 \left(3 + \frac{2}{A_{\rm esc}\bar{A}_{\rm SHA}}\right)\right]\gamma^r = 0\,.
\end{equation}
Noting that for the quadratic equation with one variable, $\Delta > 0$ can be guaranteed. So the solutions are:
\begin{equation}
    r_{\pm} = -\frac{1}{2}\pm \sqrt{\frac{25}{4}+\frac{2}{A_{\rm esc}\bar{A}_{\rm SHA}}}\,,
\end{equation}
thus Eq. (\ref{esol}) can be obtained. It is necessary to point out that here the diffusive approximation is still applied to calculate the escaping rate. However, when the MFPs of the electrons become large enough, it might not be reasonable to describe the escape of particles as diffusion processes because the lacking in efficient wave-particle resonance would make electrons propagate through the jet ballistically. Further studies involving intense MHD simulations are needed to explain the wave-particle interactions and acceleration processes in this situation.

\vspace{5mm}

\bibliography{sample701}{}

@ARTICLE{Zhang2025,
       author = {{Zhang}, B. Theodore and {Kimura}, Shigeo S. and {Murase}, Kohta},
        title = "{Microquasar jet-cocoon systems as PeVatrons}",
      journal = {arXiv e-prints},
         year = 2025,
        month = jun,
          eid = {arXiv:2506.20193},
        pages = {arXiv:2506.20193},
          doi = {10.48550/arXiv.2506.20193},
archivePrefix = {arXiv},
       eprint = {2506.20193},
 primaryClass = {astro-ph.HE},
       adsurl = {https://ui.adsabs.harvard.edu/abs/2025arXiv250620193Z}
}

@INPROCEEDINGS{LACT2024,
       author = {{Zhang}, S. and {Wang}, Y. and {Liu}, J. and {Feng}, S. and {Yang}, M. and {Geng}, L. and {2024}, Y. and {LACT group}},
        title = "{Large Array of imaging atmospheric Cherenkov Telescopes (LACT): status and future plans}",
    booktitle = {38th International Cosmic Ray Conference},
         year = 2024,
        month = sep,
          eid = {808},
        pages = {808},
       adsurl = {https://ui.adsabs.harvard.edu/abs/2024icrc.confE.808Z}
}

@ARTICLE{CTA2011,
       author = {{Actis}, M. and {Agnetta}, G. and {Aharonian}, F. and {Akhperjanian}, A. and {Aleksi{\'c}}, J. and {Aliu}, E. and {Allan}, D. and {Allekotte}, I. and {Antico}, F. and {Antonelli}, L.~A. and {Antoranz}, P. and {Aravantinos}, A. and {Arlen}, T. and {Arnaldi}, H. and {Artmann}, S. and {Asano}, K. and {Asorey}, H. and {B{\"a}hr}, J. and {Bais}, A. and {Baixeras}, C. and {Bajtlik}, S. and {Balis}, D. and {Bamba}, A. and {Barbier}, C. and {Barcel{\'o}}, M. and {Barnacka}, A. and {Barnstedt}, J. and {Barres de Almeida}, U. and {Barrio}, J.~A. and {Basso}, S. and {Bastieri}, D. and {Bauer}, C. and {Becerra}, J. and {Becherini}, Y. and {Bechtol}, K. and {Becker}, J. and {Beckmann}, V. and {Bednarek}, W. and {Behera}, B. and {Beilicke}, M. and {Belluso}, M. and {Benallou}, M. and {Benbow}, W. and {Berdugo}, J. and {Berger}, K. and {Bernardino}, T. and {Bernl{\"o}hr}, K. and {Biland}, A. and {Billotta}, S. and {Bird}, T. and {Birsin}, E. and {Bissaldi}, E. and {Blake}, S. and {Blanch}, O. and {Bobkov}, A.~A. and {Bogacz}, L. and {Bogdan}, M. and {Boisson}, C. and {Boix}, J. and {Bolmont}, J. and {Bonanno}, G. and {Bonardi}, A. and {Bonev}, T. and {Borkowski}, J. and {Botner}, O. and {Bottani}, A. and {Bourgeat}, M. and {Boutonnet}, C. and {Bouvier}, A. and {Brau-Nogu{\'e}}, S. and {Braun}, I. and {Bretz}, T. and {Briggs}, M.~S. and {Brun}, P. and {Brunetti}, L. and {Buckley}, J.~H. and {Bugaev}, V. and {B{\"u}hler}, R. and {Bulik}, T. and {Busetto}, G. and {Buson}, S. and {Byrum}, K. and {Cailles}, M. and {Cameron}, R. and {Canestrari}, R. and {Cantu}, S. and {Carmona}, E. and {Carosi}, A. and {Carr}, J. and {Carton}, P.~H. and {Casiraghi}, M. and {Castarede}, H. and {Catalano}, O. and {Cavazzani}, S. and {Cazaux}, S. and {Cerruti}, B. and {Cerruti}, M. and {Chadwick}, P.~M. and {Chiang}, J. and {Chikawa}, M. and {Cie{\'s}lar}, M. and {Ciesielska}, M. and {Cillis}, A. and {Clerc}, C. and {Colin}, P. and {Colom{\'e}}, J. and {Compin}, M. and {Conconi}, P. and {Connaughton}, V. and {Conrad}, J. and {Contreras}, J.~L. and {Coppi}, P. and {Corlier}, M. and {Corona}, P. and {Corpace}, O. and {Corti}, D. and {Cortina}, J. and {Costantini}, H. and {Cotter}, G. and {Courty}, B. and {Couturier}, S. and {Covino}, S. and {Croston}, J. and {Cusumano}, G. and {Daniel}, M.~K. and {Dazzi}, F. and {de Angelis}, A. and {de Cea Del Pozo}, E. and {de Gouveia Dal Pino}, E.~M. and {de Jager}, O. and {de La Calle P{\'e}rez}, I. and {de La Vega}, G. and {de Lotto}, B. and {de Naurois}, M. and {de O{\~n}a Wilhelmi}, E. and {de Souza}, V. and {Decerprit}, B. and {Deil}, C. and {Delagnes}, E. and {Deleglise}, G. and {Delgado}, C. and {Dettlaff}, T. and {di Paolo}, A. and {di Pierro}, F. and {D{\'\i}az}, C. and {Dick}, J. and {Dickinson}, H. and {Digel}, S.~W. and {Dimitrov}, D. and {Disset}, G. and {Djannati-Ata{\"\i}}, A. and {Doert}, M. and {Domainko}, W. and {Dorner}, D. and {Doro}, M. and {Dournaux}, J. -L. and {Dravins}, D. and {Drury}, L. and {Dubois}, F. and {Dubois}, R. and {Dubus}, G. and {Dufour}, C. and {Durand}, D. and {Dyks}, J. and {Dyrda}, M. and {Edy}, E. and {Egberts}, K. and {Eleftheriadis}, C. and {Elles}, S. and {Emmanoulopoulos}, D. and {Enomoto}, R. and {Ernenwein}, J. -P. and {Errando}, M. and {Etchegoyen}, A. and {Falcone}, A.~D. and {Farakos}, K. and {Farnier}, C. and {Federici}, S. and {Feinstein}, F. and {Ferenc}, D. and {Fillin-Martino}, E. and {Fink}, D. and {Finley}, C. and {Finley}, J.~P. and {Firpo}, R. and {Florin}, D. and {F{\"o}hr}, C. and {Fokitis}, E. and {Font}, Ll. and {Fontaine}, G. and {Fontana}, A. and {F{\"o}rster}, A. and {Fortson}, L. and {Fouque}, N. and {Fransson}, C. and {Fraser}, G.~W. and {Fresnillo}, L. and {Fruck}, C. and {Fujita}, Y. and {Fukazawa}, Y.},
        title = "{Design concepts for the Cherenkov Telescope Array CTA: an advanced facility for ground-based high-energy gamma-ray astronomy}",
      journal = {Experimental Astronomy},
         year = 2011,
        month = dec,
       volume = {32},
       number = {3},
        pages = {193-316},
          doi = {10.1007/s10686-011-9247-0},
archivePrefix = {arXiv},
       eprint = {1008.3703},
 primaryClass = {astro-ph.IM},
       adsurl = {https://ui.adsabs.harvard.edu/abs/2011ExA....32..193A}
}

@ARTICLE{Coppi1997,
       author = {{Coppi}, Paolo S. and {Aharonian}, Felix A.},
        title = "{Constraints on the Very High Energy Emissivity of the Universe from the Diffuse GeV Gamma-Ray Background}",
      journal = {\apjl},
         year = 1997,
        month = sep,
       volume = {487},
       number = {1},
        pages = {L9-L12},
          doi = {10.1086/310883},
archivePrefix = {arXiv},
       eprint = {astro-ph/9610176},
 primaryClass = {astro-ph},
       adsurl = {https://ui.adsabs.harvard.edu/abs/1997ApJ...487L...9C}
}

@ARTICLE{ASTRI-MINI22,
       author = {{Scuderi}, S. and {Giuliani}, A. and {Pareschi}, G. and {Tosti}, G. and {Catalano}, O. and {Amato}, E. and {Antonelli}, L.~A. and {Becerra Gonz{\`a}les}, J. and {Bellassai}, G. and {Bigongiari}, C. and {Biondo}, B. and {B{\"o}ttcher}, M. and {Bonanno}, G. and {Bonnoli}, G. and {Bruno}, P. and {Bulgarelli}, A. and {Canestrari}, R. and {Capalbi}, M. and {Caraveo}, P. and {Cardillo}, M. and {Conforti}, V. and {Contino}, G. and {Corpora}, M. and {Costa}, A. and {Cusumano}, G. and {D'A{\`\i}}, A. and {de Gouveia Dal Pino}, E. and {Della Ceca}, R. and {Escribano Rodriguez}, E. and {Falceta-Gon{\c{c}}alves}, D. and {Fermino}, C. and {Fiori}, M. and {Fioretti}, V. and {Fiorini}, M. and {Gallozzi}, S. and {Gargano}, C. and {Garozzo}, S. and {Germani}, S. and {Ghedina}, A. and {Gianotti}, F. and {Giarrusso}, S. and {Gimenes}, R. and {Giordano}, V. and {Grillo}, A. and {Grivel Gelly}, C. and {Impiombato}, D. and {Incardona}, F. and {Incorvaia}, S. and {Iovenitti}, S. and {La Barbera}, A. and {La Palombara}, N. and {La Parola}, V. and {Lamastra}, A. and {Lessio}, L. and {Leto}, G. and {Lo Gerfo}, F. and {Lodi}, M. and {Lombardi}, S. and {Longo}, F. and {Lucarelli}, F. and {Maccarone}, M.~C. and {Marano}, D. and {Martinetti}, E. and {Mereghetti}, S. and {Miccich{\'e}}, A. and {Millul}, R. and {Mineo}, T. and {Mollica}, D. and {Morlino}, G. and {Morselli}, A. and {Naletto}, G. and {Nicotra}, G. and {Pagliaro}, A. and {Parmiggiani}, N. and {Piano}, G. and {Pintore}, F. and {Poretti}, E. and {Olmi}, B. and {Rodeghiero}, G. and {Rodriguez Fernandez}, G. and {Romano}, P. and {Romeo}, G. and {Russo}, F. and {Sangiorgi}, P. and {Saturni}, F.~G. and {Schwarz}, J.~H. and {Sciacca}, E. and {Sironi}, G. and {Sottile}, G. and {Stamerra}, A. and {Tagliaferri}, G. and {Testa}, V. and {Umana}, G. and {Uslenghi}, M. and {Vercellone}, S. and {Zampieri}, L. and {Zanmar Sanchez}, R.},
        title = "{The ASTRI Mini-Array of Cherenkov telescopes at the Observatorio del Teide}",
      journal = {Journal of High Energy Astrophysics},
         year = 2022,
        month = aug,
       volume = {35},
        pages = {52-68},
          doi = {10.1016/j.jheap.2022.05.001},
archivePrefix = {arXiv},
       eprint = {2208.04571},
 primaryClass = {astro-ph.IM},
       adsurl = {https://ui.adsabs.harvard.edu/abs/2022JHEAp..35...52S}
}

@ARTICLE{Wang:ApJL:2025,
       author = {{Wang}, Jieshuang and {Reville}, Brian and {Aharonian}, Felix A.},
        title = "{Galactic Superaccreting X-Ray Binaries as Super-PeVatron Accelerators}",
      journal = {ApJL},
         year = 2025,
        month = aug,
       volume = {989},
       number = {2},
          eid = {L25},
        pages = {L25},
          doi = {10.3847/2041-8213/adf3a4},
archivePrefix = {arXiv},
       eprint = {2507.21048},
 primaryClass = {astro-ph.HE},
       adsurl = {https://ui.adsabs.harvard.edu/abs/2025ApJ...989L..25W}
}

@ARTICLE{Marti:ApJ:1997,
       author = {{Mart{\'\i}}, J.~M. and {M{\"u}ller}, E. and {Font}, J.~A. and {Ib{\'a}{\~n}ez}, J.~M.~Z. and {Marquina}, A.},
        title = "{Morphology and Dynamics of Relativistic Jets}",
      journal = {ApJ},
         year = 1997,
        month = apr,
       volume = {479},
       number = {1},
        pages = {151-163},
          doi = {10.1086/303842},
       adsurl = {https://ui.adsabs.harvard.edu/abs/1997ApJ...479..151M}
}

@ARTICLE{He2023,
       author = {{He}, Jia-Chun and {Sun}, Xiao-Na and {Wang}, Jie-Shuang and {Rieger}, Frank M. and {Liu}, Ruo-Yu and {Liang}, En-Wei},
        title = "{Studying X-ray spectra from large-scale jets of FR II radio galaxies: application of shear particle acceleration}",
      journal = {\mnras},
         year = 2023,
        month = nov,
       volume = {525},
       number = {4},
        pages = {5298-5310},
          doi = {10.1093/mnras/stad2542},
archivePrefix = {arXiv},
       eprint = {2308.11370},
 primaryClass = {astro-ph.HE},
       adsurl = {https://ui.adsabs.harvard.edu/abs/2023MNRAS.525.5298H}
}

@ARTICLE{Liu2021crab,
       author = {{Liu}, Ruo-Yu and {Wang}, Xiang-Yu},
        title = "{PeV Emission of the Crab Nebula: Constraints on the Proton Content in Pulsar Wind and Implications}",
      journal = {\apj},
         year = 2021,
        month = dec,
       volume = {922},
       number = {2},
          eid = {221},
        pages = {221},
          doi = {10.3847/1538-4357/ac2ba0},
archivePrefix = {arXiv},
       eprint = {2109.14148},
 primaryClass = {astro-ph.HE},
       adsurl = {https://ui.adsabs.harvard.edu/abs/2021ApJ...922..221L}
}

@ARTICLE{HESS2020,
       author = {{H.~E.~S.~S. Collaboration} and {Abdalla}, H. and {Adam}, R. and {Aharonian}, F. and {Ait Benkhali}, F. and {Ang{\"u}ner}, E.~O. and {Arakawa}, M. and {Arcaro}, C. and {Armand}, C. and {Ashkar}, H. and {Backes}, M. and {Barbosa Martins}, V. and {Barnard}, M. and {Becherini}, Y. and {Berge}, D. and {Bernl{\"o}hr}, K. and {Blackwell}, R. and {B{\"o}ttcher}, M. and {Boisson}, C. and {Bolmont}, J. and {Bonnefoy}, S. and {Bregeon}, J. and {Breuhaus}, M. and {Brun}, F. and {Brun}, P. and {Bryan}, M. and {B{\"u}chele}, M. and {Bulik}, T. and {Bylund}, T. and {Capasso}, M. and {Caroff}, S. and {Carosi}, A. and {Casanova}, S. and {Cerruti}, M. and {Chand}, T. and {Chandra}, S. and {Chen}, A. and {Colafrancesco}, S. and {Cury{\l}o}, M. and {Davids}, I.~D. and {Deil}, C. and {Devin}, J. and {deWilt}, P. and {Dirson}, L. and {Djannati-Ata{\"\i}}, A. and {Dmytriiev}, A. and {Donath}, A. and {Doroshenko}, V. and {Drury}, L. O'C. and {Dyks}, J. and {Egberts}, K. and {Emery}, G. and {Ernenwein}, J. -P. and {Eschbach}, S. and {Feijen}, K. and {Fegan}, S. and {Fiasson}, A. and {Fontaine}, G. and {Funk}, S. and {F{\"u}{\ss}ling}, M. and {Gabici}, S. and {Gallant}, Y.~A. and {Gat{\'e}}, F. and {Giavitto}, G. and {Glawion}, D. and {Glicenstein}, J.~F. and {Gottschall}, D. and {Grondin}, M. -H. and {Hahn}, J. and {Haupt}, M. and {Heinzelmann}, G. and {Henri}, G. and {Hermann}, G. and {Hinton}, J.~A. and {Hofmann}, W. and {Hoischen}, C. and {Holch}, T.~L. and {Holler}, M. and {Horns}, D. and {Huber}, D. and {Iwasaki}, H. and {Jamrozy}, M. and {Jankowsky}, D. and {Jankowsky}, F. and {Jardin-Blicq}, A. and {Jung-Richardt}, I. and {Kastendieck}, M.~A. and {Katarzy{\'n}ski}, K. and {Katsuragawa}, M. and {Katz}, U. and {Khangulyan}, D. and {Kh{\'e}lifi}, B. and {King}, J. and {Klepser}, S. and {Klu{\'z}niak}, W. and {Komin}, N. and {Kosack}, K. and {Kostunin}, D. and {Kraus}, M. and {Lamanna}, G. and {Lau}, J. and {Lemi{\`e}re}, A. and {Lemoine-Goumard}, M. and {Lenain}, J. -P. and {Leser}, E. and {Levy}, C. and {Lohse}, T. and {Lypova}, I. and {Mackey}, J. and {Majumdar}, J. and {Malyshev}, D. and {Marandon}, V. and {Marcowith}, A. and {Mares}, A. and {Mariaud}, C. and {Mart{\'\i}-Devesa}, G. and {Marx}, R. and {Maurin}, G. and {Meintjes}, P.~J. and {Mitchell}, A.~M.~W. and {Moderski}, R. and {Mohamed}, M. and {Mohrmann}, L. and {Moore}, C. and {Moulin}, E. and {Muller}, J. and {Murach}, T. and {Nakashima}, S. and {de Naurois}, M. and {Ndiyavala}, H. and {Niederwanger}, F. and {Niemiec}, J. and {Oakes}, L. and {O'Brien}, P. and {Odaka}, H. and {Ohm}, S. and {de Ona Wilhelmi}, E. and {Ostrowski}, M. and {Oya}, I. and {Panter}, M. and {Parsons}, R.~D. and {Perennes}, C. and {Petrucci}, P. -O. and {Peyaud}, B. and {Piel}, Q. and {Pita}, S. and {Poireau}, V. and {Priyana Noel}, A. and {Prokhorov}, D.~A. and {Prokoph}, H. and {P{\"u}hlhofer}, G. and {Punch}, M. and {Quirrenbach}, A. and {Raab}, S. and {Rauth}, R. and {Reimer}, A. and {Reimer}, O. and {Remy}, Q. and {Renaud}, M. and {Rieger}, F. and {Rinchiuso}, L. and {Romoli}, C. and {Rowell}, G. and {Rudak}, B. and {Ruiz-Velasco}, E. and {Sahakian}, V. and {Saito}, S. and {Sanchez}, D.~A. and {Santangelo}, A. and {Sasaki}, M. and {Schlickeiser}, R. and {Sch{\"u}ssler}, F. and {Schulz}, A. and {Schutte}, H.~M. and {Schwanke}, U. and {Schwemmer}, S. and {Seglar-Arroyo}, M. and {Senniappan}, M. and {Seyffert}, A.~S. and {Shafi}, N. and {Shiningayamwe}, K. and {Simoni}, R. and {Sinha}, A. and {Sol}, H. and {Specovius}, A. and {Spir-Jacob}, M. and {Stawarz}, {\L}. and {Steenkamp}, R. and {Stegmann}, C. and {Steppa}, C. and {Takahashi}, T. and {Tavernier}, T. and {Taylor}, A.~M. and {Terrier}, R. and {Tiziani}, D. and {Tluczykont}, M. and {Trichard}, C. and {Tsirou}, M. and {Tsuji}, N. and {Tuffs}, R.},
        title = "{Resolving acceleration to very high energies along the jet of Centaurus A}",
      journal = {\nat},
         year = 2020,
        month = jun,
       volume = {582},
       number = {7812},
        pages = {356-359},
          doi = {10.1038/s41586-020-2354-1},
archivePrefix = {arXiv},
       eprint = {2007.04823},
 primaryClass = {astro-ph.HE},
       adsurl = {https://ui.adsabs.harvard.edu/abs/2020Natur.582..356H}
}

@article{bosch2009understanding,
  title={Understanding the very-high-energy emission from microquasars},
  author={Bosch-Ramon, Valenti and Khangulyan, Dmitry},
  journal={International Journal of Modern Physics D},
  volume={18},
  number={03},
  pages={347--387},
  year={2009},
  publisher={World Scientific}
}

@article{alfaro2024ultra,
  title={Ultra-high-energy gamma-ray bubble around microquasar V4641 Sgr},
  author={Alfaro, R and Alvarez, C and Arteaga-Vel{\'a}zquez, JC and Avila Rojas, D and Ayala Solares, HA and Babu, R and Belmont-Moreno, E and Caballero-Mora, KS and Capistr{\'a}n, T and Carrami{\~n}ana, A and others},
  journal={Nature},
  volume={634},
  number={8034},
  pages={557--560},
  year={2024},
  publisher={Nature Publishing Group UK London}
}

@ARTICLE{Webb2019,
       author = {{Webb}, G.~M. and {Al-Nussirat}, S. and {Mostafavi}, P. and {Barghouty}, A.~F. and {Li}, G. and {le Roux}, J.~A. and {Zank}, G.~P.},
        title = "{Particle Acceleration by Cosmic Ray Viscosity in Radio-jet Shear Flows}",
      journal = {\apj},
         year = 2019,
        month = aug,
       volume = {881},
       number = {2},
          eid = {123},
        pages = {123},
          doi = {10.3847/1538-4357/ab2fca},
       adsurl = {https://ui.adsabs.harvard.edu/abs/2019ApJ...881..123W}
}

@ARTICLE{lhaaso2024ultrahigh,
       author = {{LHAASO Collaboration}},
        title = "{Ultrahigh-Energy Gamma-ray Emission Associated with Black Hole-Jet Systems}",
      journal = {arXiv e-prints},
         year = 2024,
        month = oct,
          eid = {arXiv:2410.08988},
        pages = {arXiv:2410.08988},
          doi = {10.48550/arXiv.2410.08988},
archivePrefix = {arXiv},
       eprint = {2410.08988},
 primaryClass = {astro-ph.HE},
       adsurl = {https://ui.adsabs.harvard.edu/abs/2024arXiv241008988L}
}

@article{markwardt1999variable,
  title={Variable-frequency quasi-periodic oscillations from the galactic microquasar GRS 1915+ 105},
  author={Markwardt, Craig B and Swank, Jean H and Taam, Ronald E},
  journal={The Astrophysical Journal},
  volume={513},
  number={1},
  pages={L37},
  year={1999},
  publisher={IOP Publishing}
}

@article{revnivtsev2002super,
  title={Super-Eddington outburst of V4641 Sgr},
  author={Revnivtsev, Mikhail and Gilfanov, Marat and Churazov, Eugene and Sunyaev, Rashid},
  journal={Astronomy \& Astrophysics},
  volume={391},
  number={3},
  pages={1013--1022},
  year={2002},
  publisher={EDP Sciences}
}

@article{orosz2001black,
  title={A Black Hole in the Superluminal Source SAX J1819. 3--2525 (V4641 Sgr)},
  author={Orosz, Jerome A and Kuulkers, Erik and van der Klis, Michiel and McClintock, Jeffrey E and Garcia, Michael R and Callanan, Paul J and Bailyn, Charles D and Jain, Raj K and Remillard, Ronald A},
  journal={The Astrophysical Journal},
  volume={555},
  number={1},
  pages={489},
  year={2001},
  publisher={IOP Publishing}
}

@article{macdonald2014black,
  title={The black hole binary V4641 Sagitarii: activity in quiescence and improved mass determinations},
  author={MacDonald, Rachel KD and Bailyn, Charles D and Buxton, Michelle and Cantrell, Andrew G and Chatterjee, Ritaban and Kennedy-Shaffer, Ross and Orosz, Jerome A and Markwardt, Craig B and Swank, Jean H},
  journal={The Astrophysical Journal},
  volume={784},
  number={1},
  pages={2},
  year={2014},
  publisher={IOP Publishing}
}

@article{hjellming2000light,
  title={Light curves and radio structure of the 1999 September transient event in V4641 Sagittarii (= XTE J1819--254= SAX J1819. 3--2525)},
  author={Hjellming, RM and Rupen, MP and Hunstead, RW and Campbell-Wilson, D and Mioduszewski, AJ and Gaensler, BM and Smith, DA and Sault, RJ and Fender, RP and Spencer, RE and others},
  journal={The Astrophysical Journal},
  volume={544},
  number={2},
  pages={977},
  year={2000},
  publisher={IOP Publishing}
}

@article{gandhi2019gaia,
  title={Gaia Data Release 2 distances and peculiar velocities for Galactic black hole transients},
  author={Gandhi, Poshak and Rao, Anjali and Johnson, Michael AC and Paice, John A and Maccarone, Thomas J},
  journal={Monthly Notices of the Royal Astronomical Society},
  volume={485},
  number={2},
  pages={2642--2655},
  year={2019},
  publisher={Oxford University Press}
}

@article{hess2024acceleration,
  title={Acceleration and transport of relativistic electrons in the jets of the microquasar SS 433},
  author={HESS Collaboration*† and Aharonian, F and Benkhali, F Ait and Aschersleben, J and Ashkar, H and Backes, M and Martins, V Barbosa and Batzofin, R and Becherini, Y and Berge, D and others},
  journal={Science},
  volume={383},
  number={6681},
  pages={402--406},
  year={2024},
  publisher={American Association for the Advancement of Science}
}

@article{ohira2024very,
  title={Very-high-energy gamma rays from cosmic rays escaping from Galactic black hole binaries},
  author={Ohira, Yutaka},
  journal={arXiv preprint arXiv:2410.22976},
  year={2024}
}

@article{neronov2025multimessenger,
  title={Multimessenger signature of cosmic rays from the microquasar V4641 Sgr propagating along a Galactic magnetic field line},
  author={Neronov, Andrii and Oikonomou, Foteini and Semikoz, Dmitri},
  journal={Physical Review D},
  volume={111},
  number={10},
  pages={103025},
  year={2025},
  publisher={APS}
}

@inproceedings{rieger2007fermi,
  title={Fermi acceleration in astrophysical jets},
  author={Rieger, Frank M and Bosch-Ramon, Valent{\'\i} and Duffy, Peter},
  booktitle={The Multi-Messenger Approach to High-Energy Gamma-Ray Sources},
  pages={119--125},
  year={2007},
  organization={Springer}
}

@article{rieger2019introduction,
  title={An introduction to particle acceleration in shearing flows},
  author={Rieger, Frank M},
  journal={Galaxies},
  volume={7},
  number={3},
  pages={78},
  year={2019},
  publisher={MDPI}
}

@article{rieger2006microscopic,
  title={A microscopic analysis of shear acceleration},
  author={Rieger, Frank M and Duffy, Peter},
  journal={The Astrophysical Journal},
  volume={652},
  number={2},
  pages={1044},
  year={2006},
  publisher={IOP Publishing}
}

@article{webb2018particle,
  title={Particle acceleration due to cosmic-ray viscosity and fluid shear in astrophysical jets},
  author={Webb, GM and Barghouty, AF and Hu, Q and Le Roux, JA},
  journal={The Astrophysical Journal},
  volume={855},
  number={1},
  pages={31},
  year={2018},
  publisher={IOP Publishing}
}

@article{wang2021particle,
  title={Particle acceleration in shearing flows: the case for large-scale jets},
  author={Wang, Jie-Shuang and Reville, Brian and Liu, Ruo-Yu and Rieger, Frank M and Aharonian, Felix A},
  journal={Monthly Notices of the Royal Astronomical Society},
  volume={505},
  number={1},
  pages={1334--1341},
  year={2021},
  publisher={Oxford University Press}
}

@article{liu2017particle,
  title={Particle acceleration in mildly relativistic shearing flows: the interplay of systematic and stochastic effects, and the origin of the extended high-energy emission in AGN jets},
  author={Liu, Ruo-Yu and Rieger, Frank M and Aharonian, Felix A},
  journal={The Astrophysical Journal},
  volume={842},
  number={1},
  pages={39},
  year={2017},
  publisher={IOP Publishing}
}

@article{laing2014systematic,
  title={Systematic properties of decelerating relativistic jets in low-luminosity radio galaxies},
  author={Laing, RA and Bridle, AH},
  journal={Monthly Notices of the Royal Astronomical Society},
  volume={437},
  number={4},
  pages={3405--3441},
  year={2014},
  publisher={Oxford University Press}
}

@incollection{gabuzda2014parsec,
  title={Parsec-Scale Jets in Active Galactic Nuclei},
  author={Gabuzda, Denise C},
  booktitle={The Formation and Disruption of Black Hole Jets},
  pages={117--148},
  year={2014},
  publisher={Springer}
}

@article{nagai2014limb,
  title={Limb-brightened jet of 3C 84 revealed by the 43 GHz very-long-baseline-array observation},
  author={Nagai, H and Haga, T and Giovannini, Gabriele and Doi, A and Orienti, M and D'Ammando, F and Kino, M and Nakamura, M and Asada, K and Hada, K and others},
  journal={The Astrophysical Journal},
  volume={785},
  number={1},
  pages={53},
  year={2014},
  publisher={IOP Publishing}
}

@article{rieger2019particle,
  title={Particle acceleration in shearing flows: efficiencies and limits},
  author={Rieger, Frank M and Duffy, Peter},
  journal={The Astrophysical Journal Letters},
  volume={886},
  number={2},
  pages={L26},
  year={2019},
  publisher={IOP Publishing}
}

@article{berezhko1981kinetic,
  title={Kinetic analysis of the charged-particle acceleration process in collisionless plasma shear flows},
  author={Berezhko, E and Krymskii, G},
  journal={Sov. Astron. Lett.(Engl. Transl.);(United States)},
  volume={7},
  number={5},
  year={1981},
  publisher={Institute of Cosmological Research and Aeronomy, Siberian Branch, USSR~…}
}

@article{earl1988cosmic,
  title={Cosmic-ray viscosity},
  author={Earl, JA and Jokipii, JR and Morfill, G},
  journal={Astrophysical Journal, Part 2-Letters (ISSN 0004-637X), vol. 331, Aug. 15, 1988, p. L91-L94.},
  volume={331},
  pages={L91--L94},
  year={1988}
}

@article{jokipii1990particle,
  title={Particle acceleration in step function shear flows-A microscopic analysis},
  author={Jokipii, JR and Morfill, GE},
  journal={Astrophysical Journal, Part 1 (ISSN 0004-637X), vol. 356, June 10, 1990, p. 255-258.},
  volume={356},
  pages={255--258},
  year={1990}
}

@article{mertsch2011new,
  title={A new analytic solution for 2nd-order Fermi acceleration},
  author={Mertsch, Philipp},
  journal={Journal of Cosmology and Astroparticle Physics},
  volume={2011},
  number={12},
  pages={010},
  year={2011},
  publisher={IOP Publishing}
}

@article{schlickeiser2002conversion,
  title={Conversion of relativistic pair energy into radiation in the jets of active galactic nuclei},
  author={Schlickeiser, R and Vainio, R and B{\"o}ttcher, M and Lerche, I and Pohl, M and Schuster, C},
  journal={Astronomy \& Astrophysics},
  volume={393},
  number={1},
  pages={69--87},
  year={2002},
  publisher={EDP Sciences}
}

@article{khangulyan2014simple,
  title={Simple analytical approximations for treatment of inverse Compton scattering of relativistic electrons in the blackbody radiation field},
  author={Khangulyan, Dmitry and Aharonian, Felix A and Kelner, Stanislav R},
  journal={The Astrophysical Journal},
  volume={783},
  number={2},
  pages={100},
  year={2014},
  publisher={IOP Publishing}
}

@book{abramowitz1948handbook,
  title={Handbook of mathematical functions with formulas, graphs, and mathematical tables},
  author={Abramowitz, Milton and Stegun, Irene A},
  volume={55},
  year={1948},
  publisher={US Government printing office}
}

@article{hillas1984origin,
  title={The origin of ultra-high-energy cosmic rays},
  author={Hillas, Anthony M},
  journal={IN: Annual review of astronomy and astrophysics. Volume 22. Palo Alto, CA, Annual Reviews, Inc., 1984, p. 425-444.},
  volume={22},
  pages={425--444},
  year={1984}
}

@article{zabalza2015naima,
  title={naima: a Python package for inference of relativistic particle energy distributions from observed nonthermal spectra},
  author={Zabalza, V{\'\i}ctor},
  journal={arXiv preprint arXiv:1509.03319},
  year={2015}
}

@article{dermer2009high,
  title={High energy radiation from black holes: gamma rays, cosmic rays, and neutrinos},
  author={Dermer, Charles D and Menon, Govind},
  year={2009},
  publisher={Princeton university press}
}

@article{salvesen2020origin,
  title={Origin of spin--orbit misalignments: the microblazar V4641 Sgr},
  author={Salvesen, Greg and Pokawanvit, Supavit},
  journal={Monthly Notices of the Royal Astronomical Society},
  volume={495},
  number={2},
  pages={2179--2204},
  year={2020},
  publisher={Oxford University Press}
}

@article{aharonian2010angular,
  title={Angular, spectral, and time distributions of highest energy protons and associated secondary gamma rays and neutrinos propagating through extragalactic magnetic and radiation fields},
  author={Aharonian, FA and Kelner, SR and Prosekin, A Yu},
  journal={Physical Review D—Particles, Fields, Gravitation, and Cosmology},
  volume={82},
  number={4},
  pages={043002},
  year={2010},
  publisher={APS}
}

@article{suzuki2025detection,
  title={Detection of extended X-ray emission around the PeVatron microquasar V4641 Sgr with XRISM},
  author={Suzuki, Hiromasa and Tsuji, Naomi and Kanemaru, Yoshiaki and Shidatsu, Megumi and Olivera-Nieto, Laura and Safi-Harb, Samar and Kimura, Shigeo S and de la Fuente, Eduardo and Casanova, Sabrina and Mori, Kaya and others},
  journal={The Astrophysical Journal Letters},
  volume={978},
  number={2},
  pages={L20},
  year={2025},
  publisher={IOP Publishing}
}

@article{demidem2020particle,
  title={Particle acceleration in relativistic turbulence: A theoretical appraisal},
  author={Demidem, Camilia and Lemoine, Martin and Casse, Fabien},
  journal={Physical Review D},
  volume={102},
  number={2},
  pages={023003},
  year={2020},
  publisher={APS}
}

@article{jaekel1992fokker,
  title={The Fokker-Planck coefficients of cosmic ray transport in random electromagnetic fields},
  author={Jaekel, U and Schlickeiser, R},
  journal={Journal of Physics G: Nuclear and Particle Physics},
  volume={18},
  number={6},
  pages={1089},
  year={1992},
  publisher={IOP Publishing}
}

@ARTICLE{WangJS2024,
       author = {{Wang}, Jie-Shuang and {Reville}, Brian and {Rieger}, Frank M. and {Aharonian}, Felix A.},
        title = "{Acceleration of Ultra-high-energy Cosmic Rays in the Kiloparsec-scale Jets of Nearby Radio Galaxies}",
      journal = {\apjl},
         year = 2024,
        month = dec,
       volume = {977},
       number = {1},
          eid = {L20},
        pages = {L20},
          doi = {10.3847/2041-8213/ad9589},
archivePrefix = {arXiv},
       eprint = {2411.16674},
 primaryClass = {astro-ph.HE},
       adsurl = {https://ui.adsabs.harvard.edu/abs/2024ApJ...977L..20W}
}

@ARTICLE{WangJS2023,
       author = {{Wang}, Jie-Shuang and {Reville}, Brian and {Mizuno}, Yosuke and {Rieger}, Frank M. and {Aharonian}, Felix A.},
        title = "{Particle acceleration in shearing flows: the self-generation of turbulent spine-sheath structures in relativistic magnetohydrodynamic jet simulations}",
      journal = {\mnras},
         year = 2023,
        month = feb,
       volume = {519},
       number = {2},
        pages = {1872-1880},
          doi = {10.1093/mnras/stac3616},
archivePrefix = {arXiv},
       eprint = {2212.03226},
 primaryClass = {astro-ph.HE},
       adsurl = {https://ui.adsabs.harvard.edu/abs/2023MNRAS.519.1872W}
}

@article{gallo2013v4641,
  title={V4641 Sgr: a candidate precessing microblazar},
  author={Gallo, Elena and Plotkin, Richard M and Jonker, Peter G},
  journal={Monthly Notices of the Royal Astronomical Society: Letters},
  volume={438},
  number={1},
  pages={L41--L45},
  year={2014},
  publisher={The Royal Astronomical Society}}

@article{zhao2025upper,
  title={Upper limits on the gamma-ray emission from the microquasar V4641 Sgr},
  author={Zhao, Zihao and Li, Jian and Torres, Diego F},
  journal={arXiv preprint arXiv:2503.16844},
  year={2025}
}

@article{blandford1977electromagnetic,
  title={Electromagnetic extraction of energy from Kerr black holes},
  author={Blandford, Roger D and Znajek, Roman L},
  journal={Monthly Notices of the Royal Astronomical Society},
  volume={179},
  number={3},
  pages={433--456},
  year={1977},
  publisher={Oxford University Press Oxford, UK}
}

@article{zdziarski2023evidence,
  title={Evidence for a Black Hole Spin--Orbit Misalignment in the X-Ray Binary Cyg X-1},
  author={Zdziarski, Andrzej A and Veledina, Alexandra and Szanecki, Micha{\l} and Green, David A and Bright, Joe S and Williams, David RA},
  journal={The Astrophysical Journal Letters},
  volume={951},
  number={2},
  pages={L45},
  year={2023},
  publisher={IOP Publishing}
}

@article{poutanen2022black,
  title={Black hole spin--orbit misalignment in the x-ray binary MAXI J1820+ 070},
  author={Poutanen, Juri and Veledina, Alexandra and Berdyugin, Andrei V and Berdyugina, Svetlana V and Jermak, Helen and Jonker, Peter G and Kajava, Jari JE and Kosenkov, Ilia A and Kravtsov, Vadim and Piirola, Vilppu and others},
  journal={Science},
  volume={375},
  number={6583},
  pages={874--876},
  year={2022},
  publisher={American Association for the Advancement of Science}
}

@article{fragos2010black,
  title={Black hole spin--orbit misalignment in galactic x-ray binaries},
  author={Fragos, T and Tremmel, M and Rantsiou, E and Belczynski, K},
  journal={The Astrophysical Journal Letters},
  volume={719},
  number={1},
  pages={L79},
  year={2010},
  publisher={IOP Publishing}
}

@book{kulsrud2020plasma,
  title={Plasma physics for astrophysics},
  author={Kulsrud, Russell M},
  year={2020},
  publisher={Princeton University Press}
}

@article{tetarenko2016watchdog,
  title={WATCHDOG: a comprehensive all-sky database of galactic black hole X-ray binaries},
  author={Tetarenko, BE and Sivakoff, GR and Heinke, CO and Gladstone, JC},
  journal={The Astrophysical Journal Supplement Series},
  volume={222},
  number={2},
  pages={15},
  year={2016},
  publisher={IOP Publishing}
}

@article{uemura2004outburst,
  title={Outburst and post-outburst active phase of the black hole X-ray binary V4641 Sagittarii in 2002},
  author={Uemura, Makoto and Kato, Taichi and Ishioka, Ryoko and Tanabe, Kenji and Torii, Ken’ichi and Santallo, Roland and Monard, Berto and Markwardt, Craig B and Swank, Jean H and Sault, Robert J and others},
  journal={Publications of the Astronomical Society of Japan},
  volume={56},
  number={sp1},
  pages={S61--S75},
  year={2004},
  publisher={Oxford Universtiy Press Oxford, UK}
}

@article{buxton2003optical,
  title={Optical Outburst of V4641 (= SAX J1819. 3-2525)},
  author={Buxton, M and Maitra, D and Bailyn, C and Jeanty, L and Gonzalez, D},
  journal={The Astronomer's Telegram},
  volume={170},
  pages={1},
  year={2003}
}

@article{swank2004x,
  title={X-ray Reappearance of V4641 Sgr},
  author={Swank, Jean},
  journal={The Astronomer's Telegram},
  volume={295},
  pages={1},
  year={2004}
}

@article{yoshii2015maxi,
  title={MAXI/GSC detection of renewed activity of the black hole candidate V4641 Sgr},
  author={Yoshii, T and Negoro, H and Ueno, S and Tomida, H and Nakahira, S and Kimura, M and Ishikawa, M and Nakagawa, YE and Mihara, T and Sugizaki, M and others},
  journal={The Astronomer's Telegram},
  volume={7858},
  pages={1},
  year={2015}
}

@article{cackett2007swift,
  title={Swift Observations of Rapid X-ray Flux Variations in V4641 Sgr},
  author={Cackett, EM and Miller, JM},
  journal={The Astronomer's Telegram},
  volume={1135},
  pages={1},
  year={2007}
}

@article{yamaoka2008swift,
  title={Swift XRT observations of the black hole binary V4641 Sgr},
  author={Yamaoka, Kazutaka and Homan, Jeroen and Uemura, Makoto},
  journal={The Astronomer's Telegram},
  volume={1796},
  pages={1},
  year={2008}
}

@article{yamaoka2010rxte,
  title={RXTE/PCA and Swift/XRT detect an outburst of the black hole binary V4641 Sgr},
  author={Yamaoka, Kazutaka and Nakahira, Satoshi},
  journal={The Astronomer's Telegram},
  volume={2785},
  pages={1},
  year={2010}
}

@article{tachibana2014maxi,
  title={MAXI/GSC detection of a renewed X-ray activity of the black hole candidate V4641 Sgr},
  author={Tachibana, Y and Takagi, T and Serino, M and Morii, M and Nakahira, S and Negoro, H and Ueno, S and Tomida, H and Kimura, M and Ishikawa, M and others},
  journal={The Astronomer's Telegram},
  volume={5803},
  pages={1},
  year={2014}
}

@article{amato2009kinetic,
  title={A kinetic approach to cosmic-ray-induced streaming instability at supernova shocks},
  author={Amato, Elena and Blasi, Pasquale},
  journal={Monthly Notices of the Royal Astronomical Society},
  volume={392},
  number={4},
  pages={1591--1600},
  year={2009},
  publisher={Blackwell Publishing Ltd Oxford, UK}
}
\bibliographystyle{aasjournalv7}

\begin{figure}[h!]
\centering
\includegraphics[scale=0.7]{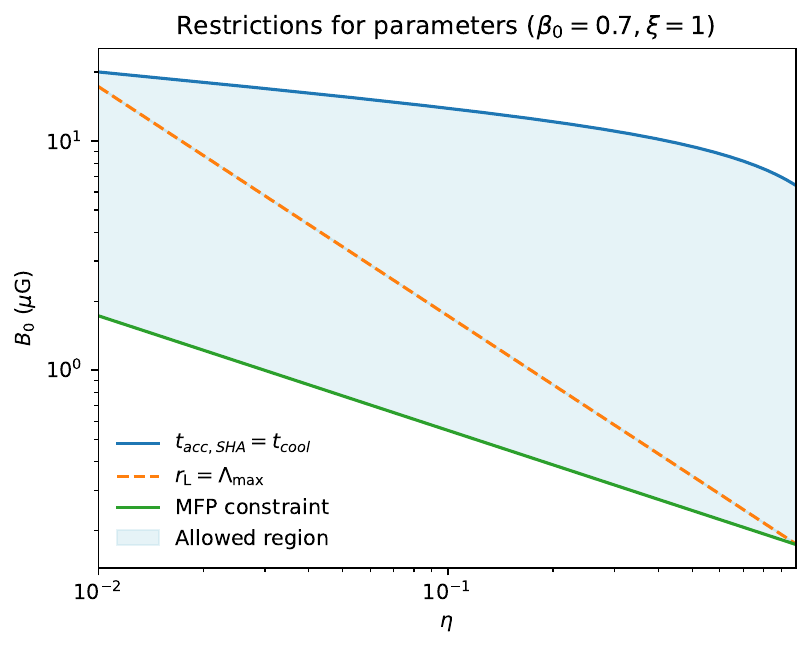}
\caption{$\eta-B_0$ planes which show the potential parameter sets under two confinements: $c\tau_{\rm sc} \le R_{\rm jet}$ and $t_{\rm acc,\rm SHA} \le t_{\rm c}$, with $\beta_0$ = 0.7, q=5/3, $R_{\rm jet}$ = 5 pc and $\gamma$ = 0.8 ${\rm PeV}/\left({m_{\rm e} c^2}\right)$. The orange dashed line indicates the limit for the 1st-order resonance and the shaded area represents the allowed parameter regions when $\xi=1$. 
\label{fig:confine}}
\end{figure}

\begin{table}[htbp]
\centering
\caption{Best-fit parameters from $\chi^2$ tests\label{tab:paras1}}
1\begin{tabular}{lcccc}
\toprule
 & \multicolumn{4}{c}{Jet radius $R_{\rm jet}$} \\
\cmidrule(lr){2-5}
Parameters & 5 pc & 3 pc & 1 pc & 0.5 pc \\ 
\midrule
$\log B_0$ ($\mu$G) 
& $-0.32^{+0.20}_{-0.35}$ & $-0.02^{+0.13}_{-0.40}$ & $0.49^{+0.07}_{-0.45}$ & $0.73^{+0.10}_{-0.40}$  \\

$\eta$ ($R_{\rm shear}/R_{\rm jet}$)  
& $0.38^{+0.39}_{-0.13}$ & $0.29^{+0.43}_{-0.07}$ & $0.28^{+0.46}_{-0.05}$ & $0.30^{+0.44}_{-0.06}$ \\

$\beta_0$ (Spine velocity)  
& $0.70^{+0.21}_{-0.08}$ & $0.64^{+0.25}_{-0.05}$ & $0.62^{+0.28}_{-0.02}$ & $0.65^{+0.25}_{-0.03}$ \\

$\log N_{\rm tot}$ (Norm.) 
& $46.90^{+0.72}_{-1.02}$ & $46.42^{+1.47}_{-0.43}$ & $46.59^{+1.95}_{-0.49}$ & $46.42^{+2.17}_{-0.25}$   \\

$\log \rm L_{\rm kin, e}$ (Electrons' kinetic luminosity, erg/s) 
& $37.29^{+1.15}_{-0.44}$ & $37.04^{+1.30}_{-0.23}$ & $37.03^{+1.41}_{-0.19}$ & $37.06^{+1.45}_{-0.17}$   \\

$\chi^2/$d.o.f. & 13.5/12 & 13.5/12 & 13.4/12 & 13.5/12  \\
\bottomrule
\end{tabular}

\vspace{0.2cm}
{\small Notes: Parameters derived from $\chi^2$ minimization across different jet radius scales.}
\end{table}

\begin{figure}[h!]
    \centering
    \includegraphics[scale=0.6]{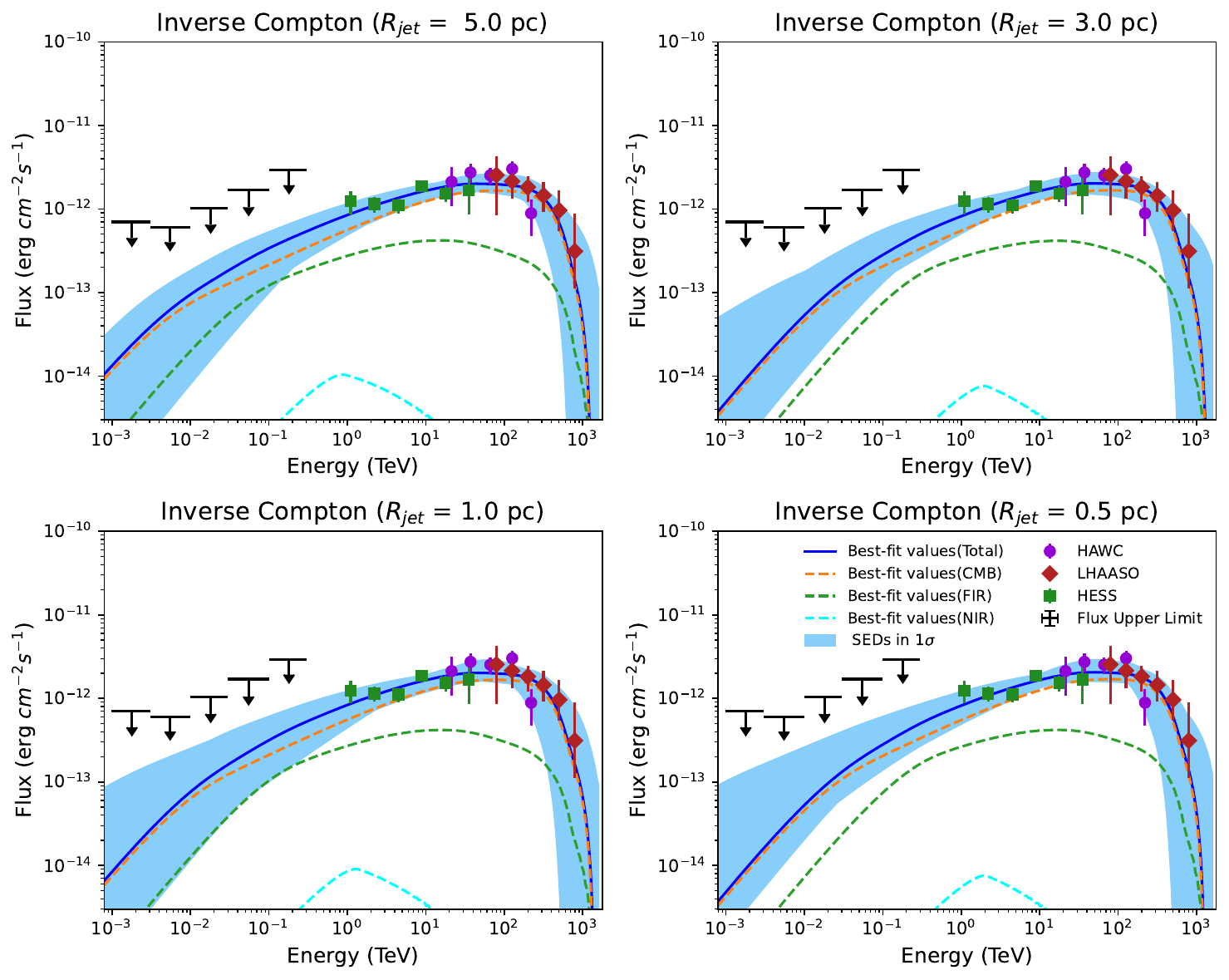}
    \caption{The fitting results for the UHE spectrum of V4641 Sgr, with data points from HAWC \citep{alfaro2024ultra}, LHAASO \citep{lhaaso2024ultrahigh} and H.E.S.S. \citep{neronov2025multimessenger}. The upper limits in 1 - 300 GeV are extracted from \citealt{zhao2025upper}. The blue solid lines indicate the spectral energy distributions (SEDs) of total IC emission from three photon fields and the lightblue regions are composed of possible fitting results in 1$\sigma$ range of further $\chi^2$ tests.}
    \label{fig:IC}
\end{figure}

\begin{figure}[h!]
    \centering
    \includegraphics[width=\textwidth]{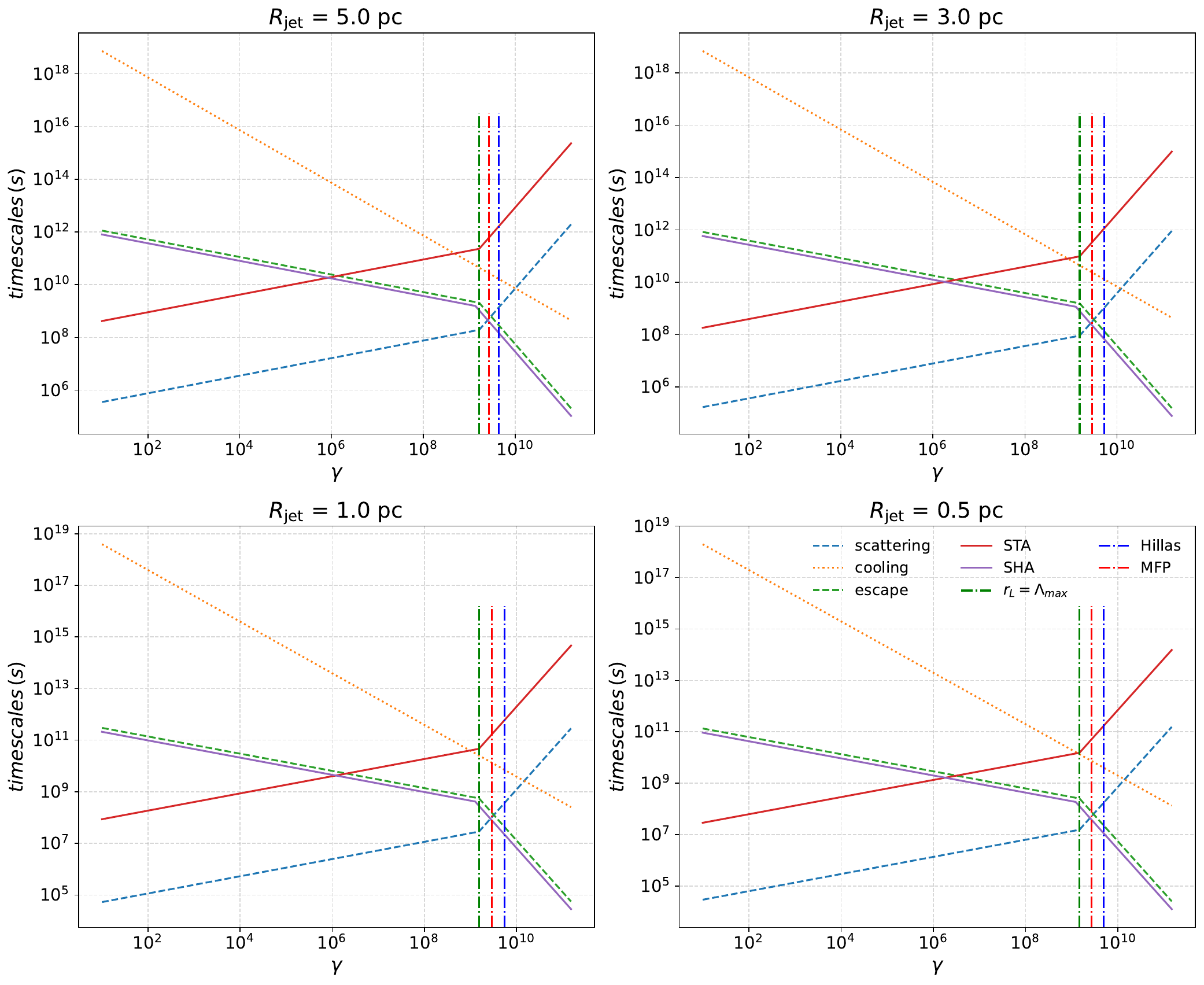}
    \caption{Timescales for different processes with the best-fit parameters in \ref{subsec:UHE}, derived from equation (\ref{e12}), (\ref{e8}), (\ref{e17}), (\ref{e20}) and (\ref{e21}). The red solid line represents the timescale for stochastic acceleration (STA) and the purple one represents the one for shear acceleration (SHA); The orange dotted line shows the timescale for radiative cooling and the blue dashed line is the scattering timescale for electrons in the jet; The dash-dotted lines indicate the extra limitations caused by the Hillas criterion ($\gamma_{\rm Hillas}$) and the particle mean free path (MFP) ($\gamma_{\rm MFP}$). }
    \label{fig:times}
\end{figure}

\begin{figure}[h!]
    \centering
    \includegraphics[scale=0.6]{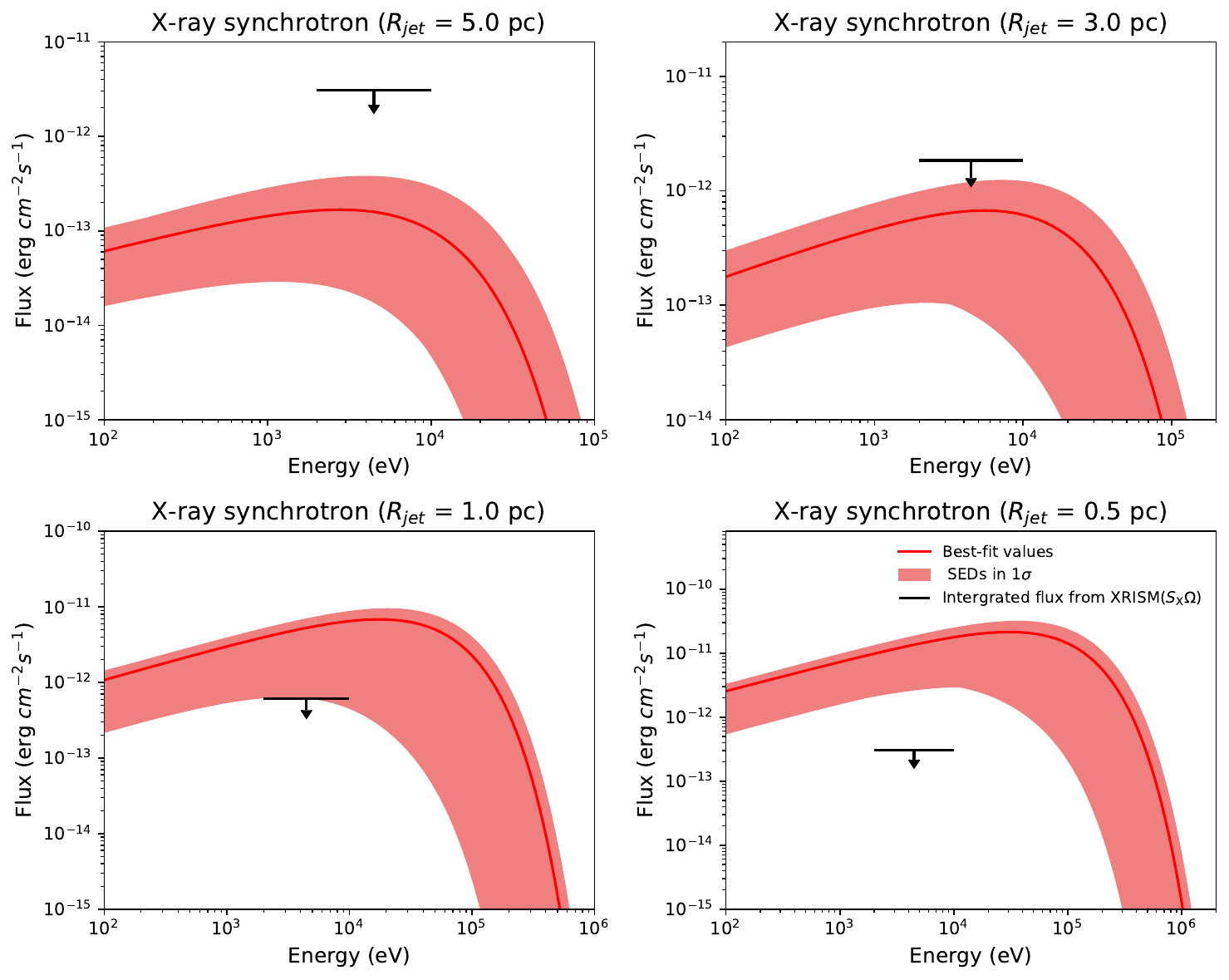}
    \caption{The synchrotron emission produced by the same electron population responsible for the UHE gamma-ray emission. The red solid lines represent the emission under the best-fit parameters and the reddish regions are composed of predicted emission with possible spectral energy distributions (SEDs) in 1$\sigma$ range of $\chi^2$ tests, which is coherent with the ones shown in Figure \ref{fig:IC}. The black solid lines represent the total integrated flux inferred from the surface brightness obtained by XRISM \citep{suzuki2025detection} from 2 to 10 keV.}
    \label{fig:SYN}
\end{figure}

\begin{figure}[h!]
    \centering
    \includegraphics[scale=0.6]{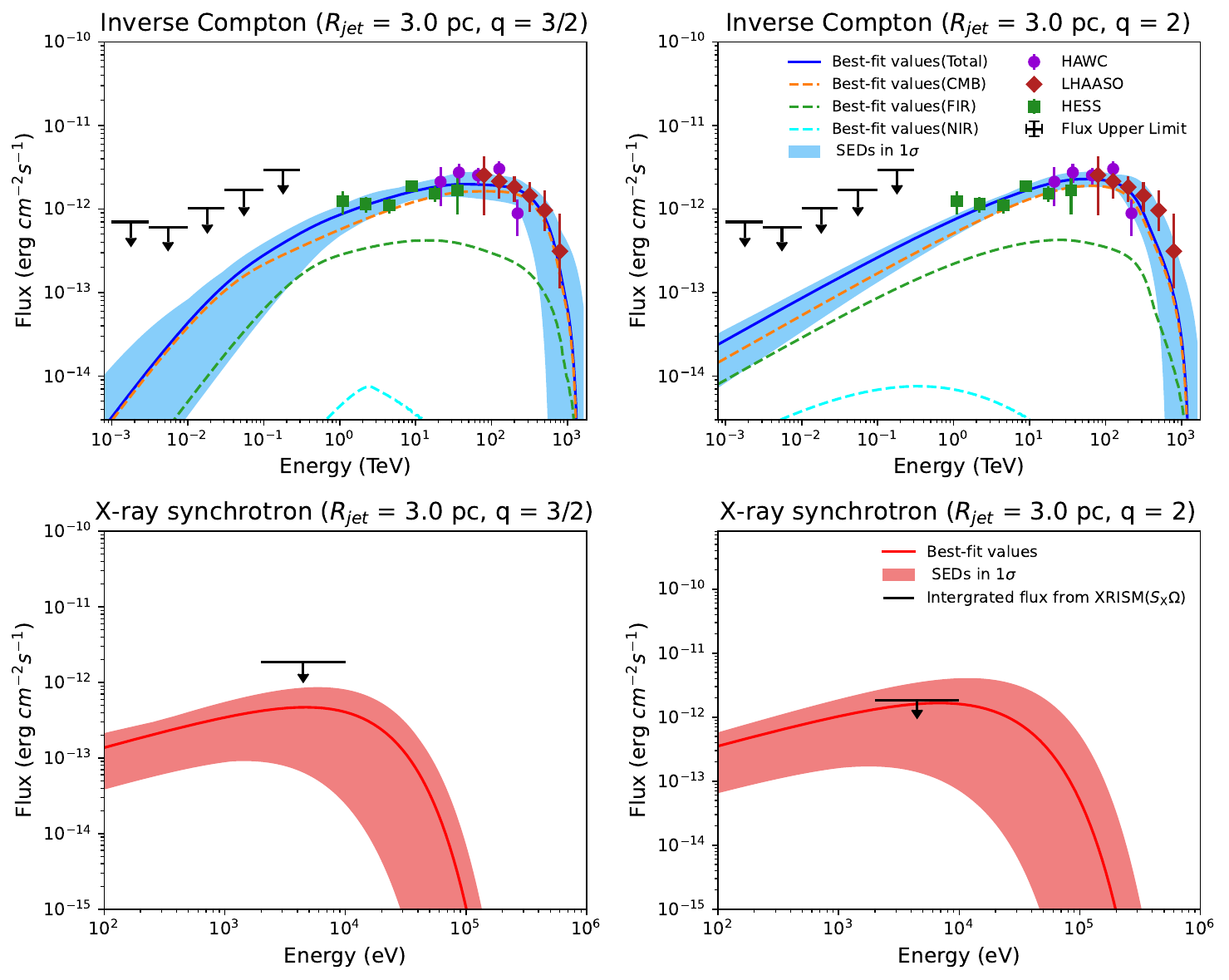}
    \caption{The fitting results for the UHE spectrum and predicted X-ray synchrotron flux, with different types of turbulence. The left-side panels show results from the Kraichnan-type ($q$ = $\rm 3/2$) while the right-side panels show that from the hard-sphere limit ($q$ = $\rm 2$). Both of the two cases can explain the observations, with only minor impacts on crucial parameters.   }
    \label{fig:VariableQ}
\end{figure}

\begin{table}[htbp]
\centering
\caption{Best-fit parameters from $\chi^2$ tests, with $R_{\rm jet}$ = 3 pc\label{tab:paras2}}
\begin{tabular}{lcc}
\toprule
 & \multicolumn{2}{c}{Type of turbulence} \\
\cmidrule(lr){2-3}
Parameters & $q = 3/2$ & $q = 2$  \\ 
\midrule
$\log B_0$ ($\mu$G) 
& $-0.09^{+0.10}_{-0.36}$ & $0.16^{+0.14}_{-0.49}$   \\

$\eta$ ($R_{\rm shear}/R_{\rm jet}$)  
& $0.42^{+0.31}_{-0.20}$ & $0.11^{+0.21}_{-0.01}$  \\

$\beta_0$ (Spine velocity)  
& $0.78^{+0.15}_{-0.10}$ & $0.38^{+0.25}_{-0.01}$  \\

$\log N_{\rm tot}$ (Norm.) 
& $46.51^{+0.84}_{-0.76}$ & $51.67^{+0.50}_{-1.41}$    \\

$\log \rm L_{\rm kin, e}$ (Electrons' kinetic luminosity, erg/s) 
& $37.49^{+1.14}_{-0.53}$ & $36.90^{+0.57}_{-0.33}$ \\

$\chi^2/$d.o.f. & 13.6/12 & 13.4/12   \\
\bottomrule
\end{tabular}

\end{table}

\end{document}